\documentstyle[epsf,aps]{revtex}

\begin{document}
\title{Comparison of advanced gravitational-wave detectors}
\author{Gregory~M.~Harry\footnote{Current address: LIGO Project,
        Massachusetts Institute of Technology, Room~NW17-161, 175~Albany
        Street, Cambridge, Massachusetts, 02139}}
\address{Department of Physics, Syracuse University, Syracuse, New York
         13244-1130}
\author{Janet~L.~Houser\footnote{Permanent address: Harvard-Smithsonian
        Center for Astrophysics, 60 Garden Street, Cambridge, Massachusetts, 02138}}
\address{LIGO Visiting Scientist, Massachusetts Institute of Technology,
         Room NW17-161, 175~Albany Street, Cambridge, Massachusetts, 02139}
\author{Kenneth~A.~Strain}
\address{Department of Physics and Astronomy, University of Glasgow,
        Glasgow G12 8QQ, Scotland, United Kingdom}
\date{\today}
\maketitle

\begin{abstract}
We compare two advanced designs for gravitational-wave antennas in
terms of their ability to detect two possible gravitational wave
sources. Spherical, resonant mass antennas and interferometers
incorporating resonant sideband extraction (RSE) were modeled using 
experimentally measurable parameters.  The signal-to-noise ratio of
each detector for a binary neutron star system and a rapidly rotating
stellar core were calculated.  For a range of plausible parameters we
found that the advanced LIGO interferometer incorporating RSE gave
higher signal-to-noise ratios than a spherical detector resonant at the
same frequency for both sources.  Spheres were found to be sensitive to
these sources at distances beyond our galaxy. Interferometers were
sensitive to these sources at far enough distances that several events
per year would be expected.
\end{abstract}


\section{Introduction}

The experimental effort to detect gravitational radiation has advanced
substantially since its beginnings in the early 1960's~\cite{Weber}.
Two different techniques, resonant-mass antennas and interferometers,
have been developed over the years, but in the last decade construction
has begun on long baseline interferometers designed for very high
sensitivity.

In previous work~\cite{Harry} the  sensitivity of these two
technologies were compared using models of potential sources of
gravitational waves. In that work, different diameter spherical,
resonant-mass antennas~\cite{Merkowitz} were compared with the expected
sensitivity of initial LIGO~\cite{Abramovici} because it was plausible
that both detectors could be in operation in the early years of the 2000
decade.  Generally, LIGO was found to be more sensitive to these
sources, especially the inspiraling binary neutron stars.  However, at
higher frequencies, spheres were shown to provide extra sensitivity
within a restricted bandwidth.  These higher frequencies, about 700~Hz
to 5000~Hz, are where the gravitational waves from binary neutron star
coalescence, rapidly rotating stellar cores, and other
sources~\cite{Thorne87} are found.

Although gravitational wave sources occur at many different frequencies
and amplitudes,  binary neutron star inspirals emit gravitational waves
at frequencies accessible to Earth-based interferometers, a few to a
few hundred Hertz. Neutron star binaries also have an amplitude and
event rate that makes it plausible that advanced interferometers will
detect one or more a year~\cite{KalogeraNarayan}.  At these
frequencies, the neutron stars are many times their own radius apart
from each other and act as point masses.  This makes analytically
predicting the waveform possible, but also means that the details of
the neutron star composition (equation of state, radius, magnetic
field, etc.) will not effect the waveform. When compact bodies such as
neutron stars are close enough together that the gravitational
radiation being given off carries information about their structure,
the frequency is higher (typically above 700~Hz) and the amplitude is
lower.  This is true of the internal motion of compact bodies as well,
such as the core collapse of a supernova. Determining astrophysically
interesting parameters from these higher frequency waveforms will be a
major goal of gravitational wave astronomy once the first detections
have occurred and may require detectors specialized for higher
frequency response.

Progress with interferometers and delays with spherical antennas make
it more relevant now to compare spheres and a more advanced
interferometer. One possible upgrade of LIGO would include an
additional mirror at the output port, which allows for signal
recycling~\cite{Meers} or resonant-sideband extraction
(RSE)~\cite{RSE}.  These techniques allow the frequency of peak
response to be selected over a wide range and the bandwidth of the
response to be controlled.  Such an upgrade would allow the advanced
interferometer to operate with a similar strain spectrum as a spherical,
resonant-mass antenna and have more sensitivity than other
interferometer configurations at higher frequencies.

Here we compare the sensitivity of advanced LIGO, including RSE, with
that of eight possible spherical antennas with different diameters to
determine the effectiveness of each technology as a detector of high
frequency gravitational radiation. The frequency of peak sensitivity
and the bandwidth for the interferometer was chosen to correspond as
closely as possible to the lowest quadrupole resonance of each of the
spheres. Different peak--sensitivity frequencies for the interferometer
are obtained by varying the position of the signal-extraction mirror
(by a fraction of the wavelength of the laser light). The fractional
bandwidth of a sphere is determined by the choice of transducer, and is
typically below 20\%. To reproduce this narrow bandwidth, the
transmittance of the interferometer's signal-extraction mirror has been
chosen to be relatively low. Once the interferometer has been matched
to a given sphere, the sensitivity of the sphere and interferometer to
different, high frequency, sources can be compared.

The signal-to-noise ratio for both the spheres and the interferometer
was computed using the numerically simulated relativistic waveforms of
two different gravitational wave (burst) sources: (1) the inspiral and
eventual coalescence of a binary neutron star system, and (2) a rapidly
rotating stellar core undergoing a dynamical
instability~\cite{Houser2000,Houser00a}. Both of these sources are
predicted to have high enough event rates that they could be detected
by interferometers within the next ten
years~\cite{Abramovici,Thorne95,P91,KalogeraNarayan,Thorne87}

There are many other sources of gravitational waves that could provide
interesting physical and/or astrophysical information.  The stochastic
background of gravitational radiation depends on conditions at the
earliest times in the universe~\cite{stochastic} and their detection
would shed new light on cosmology.  Searches for scalar radiation would
allow for tests of gravity beyond the prediction of general relativity.
Both interferometers and spheres may play a role in these experiments,
but we did not consider these sources in this work.

\section{Method}

The spherical antennas were modeled using the same method as described
in the previous paper~\cite{Harry}.  The signal-to-noise ratio density
was calculated for each sphere using the method of Price~\cite{Price}
as extended by Stevenson~\cite{Stevenson}.  This involved calculating
the signal-to-noise ratio density from the strain spectrum of each
sphere.  The parameters that entered the strain spectrum were chosen
based, as much as possible, on optimistic extrapolations from values
demonstrated in operating detectors.

\begin{figure}
\caption{Schematic drawing of a spherical, resonant mass,
gravitational wave detector with a three-mode transducer attached.  The
sphere mass, $m_s$, is connected to mechanical ground, here the center
of mass of the sphere.  The gravitational wave acts as a force, $F$,
between the sphere mass and ground. Each of the complex spring
constants, $k_s$, $k_1$, and $k_2$, includes dissipation which gives
rise to thermal noise. The transducer mass is next to a superconducting
pick up coil which stores a persistent current.  This current is
shunted to the input coil of the SQUID in proportion to the motion of
the mass. The SQUID amplifies the signal and also serves as a source of
wideband noise.}
\begin{center}
\epsfxsize=12cm \leavevmode \epsfbox{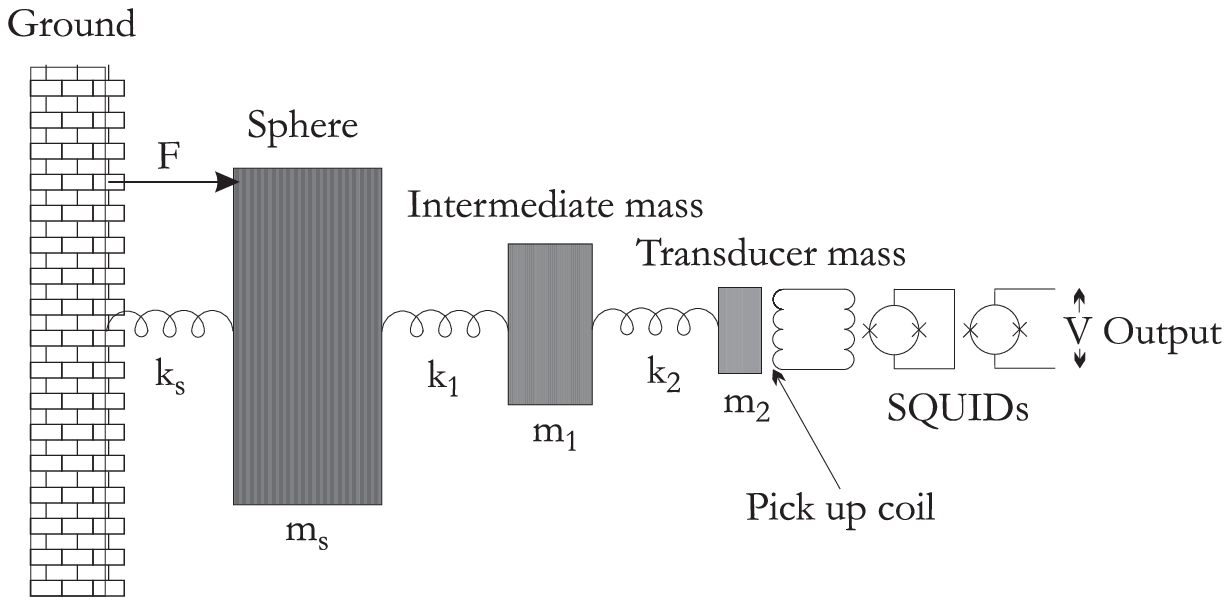}
\end{center}
\label{fig:sphere}
\end{figure}

Eight aluminum spheres were modeled with diameters of 3.25~m, 2.75~m,
2.35~m, 2.00~m, 1.70~m, 1.45~m, 1.25~m, and 1.05~m.  The 3.25~m sphere
weighs 50~tonnes and is the largest solid sphere which can reasonably
be manufactured and transported.  It may be possible to get larger
radii spheres, and hence lower resonance frequencies, by building
hollow spheres~\cite{Frossati} but we did not consider these detectors.
A lower frequency, hollow sphere will have advantages with lower
frequency sources, especially the inspiral phase of the binary neutron
star signal, but not with the higher frequency sources we considered
here. Each sphere was modeled as having six, three-mode inductive
transducers arranged in the TIGA geometry~\cite{Merkowitz} which allows
for omnidirectional sensitivity. The masses of the intermediate mass
and the final transducer mass (see Fig.~\ref{fig:sphere}) were chosen
to give a fractional bandwidth as large a possible for each sphere. The
transducer was modeled as having a dual, superconducting quantum
interference device (SQUID) as the first stage amplifier.  SQUID
amplifiers are currently in use on bar detectors~\cite{Johnson,auriga}.
A three-mode transducer is being developed for use on the Allegro
antenna~\cite{McHugh} and one has been successfully demonstrated on a
test antenna~\cite{Folkner}.

The limiting noise in a spherical resonant mass detector comes
from two sources; amplifier noise and thermal noise.  The
amplifier noise, which comes primarily from the sensing SQUID,
can be separated into additive velocity noise and force noise. The
additive velocity noise can be written as~\cite{Harry}
\begin{equation}
S_u \left(f\right) = N_n \hbar \, 2 \pi f_0/r_n,
\end{equation}
where $N_n$ is the noise number of the SQUID, $\hbar$ is Plank's
constant, $f_0$ is the resonant frequency of the sphere, and $r_n$
is the noise resistance of the transducer. We assumed that the
SQUID had quantum-limited noise, and hence a noise number equal to
one. A quantum-limited SQUID suitable for use in a gravitational
wave transducer has not yet been demonstrated, although the
quantum limit has been reached in SQUIDs with low input
impedance~\cite{Awschalom}.  A noise number of 24 has been
reached at 100~mK in a suitable SQUID when cooled on its
own~\cite{Jin}. This noise, however, was found to increase
significantly when placed in a transducer~\cite{Harry2000}.

\begin{figure}
\caption{Strain spectra for a 3.25~m diameter spherical, resonant mass antenna including the
components of the noise.  The dashed line shows the thermal noise at
50~mK, the dotted line shows the velocity (forward action) amplifier
noise from the quantum limited SQUID, and the dashed-dotted line shows
the force (back-action) amplifier noise from the same SQUID.  The solid
line is the total noise from the spherical antenna.}
\begin{center}
\epsfxsize=16cm \leavevmode \epsfbox{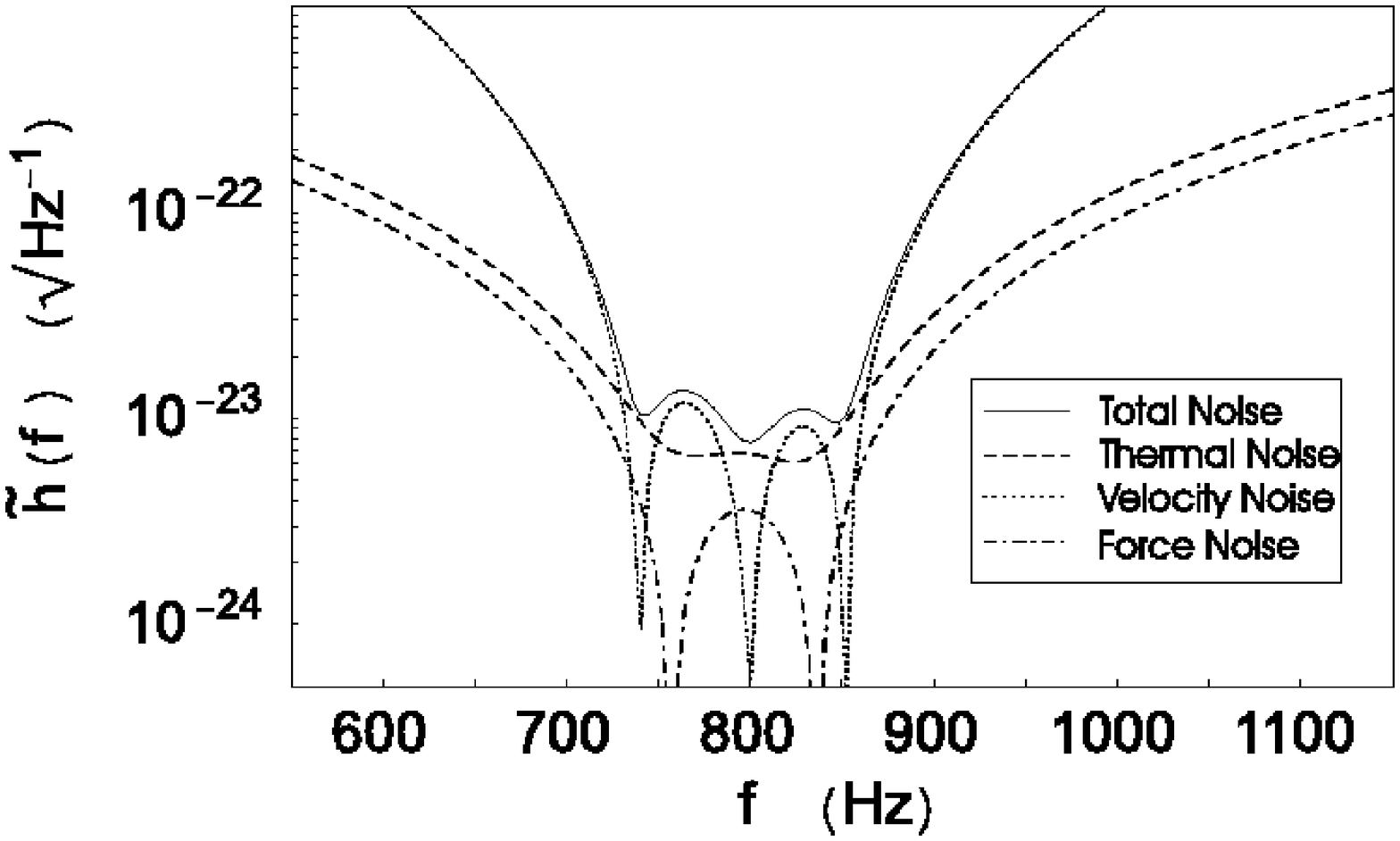}
\end{center}
\label{fig:sphereparts}
\end{figure}

The noise resistance was calculated from
\begin{equation}
r_n = \kappa A_{\mathrm{coil}}/\left(4 \pi f_{0}\right),
\end{equation}
where $\kappa$ is an experimentally determined spring constant density
given in Table~\ref{table:sphere} and $A_{\mathrm{coil}} \left(= \pi
\left(d_{c}/2\right)^{2}\right)$ is the area of the pick-up coils.
The noise resistance limits the bandwidth of a sphere with three-mode 
transducers (when the masses are properly chosen) according to~\cite{Price}
\begin{equation}
\delta_{\mathrm{BW}} \sim \left(r_n/\left(2 \pi f_{0} m_s\right)\right)^{1/5},
\end{equation}
where $m_s$ is the effective mass of the sphere for impedance
calculations. This effective mass is given
by~\cite{Stevenson,Zhou}
\begin{equation}
m_s = 5/6 \, \chi \, m_p
\end{equation}
where $\chi = 0.301$~\cite{Zhou}, $m_p$ is the physical mass of the
sphere, and the factor $5/6$ is appropriate for six transducers in the
TIGA arrangement~\cite{Stevenson}.  The amplifier velocity noise is shown 
in Fig.~\ref{fig:sphereparts} graphed as a strain spectrum.

The force noise from the SQUID is ultimately detected as motion, and
therefore must be converted by the mechanical transfer function of the
antenna.  The output noise can be written~\cite{Harry}
\begin{equation}
S_{f,\, \mathrm{out}}\left(f\right)= N_n \hbar \, 2 \pi f_0 r_n
|\,y_{22}\left(f\right)|^2,
\end{equation}
where the admittance matrix element $y_{22}\left(f\right)$ for a
sphere with three-mode transducers can be written
\begin{eqnarray}
y_{22}\left(f\right)&=& - 2 \pi f i \left( c_1 + c_2 + c_s - 4 \pi^2
f^2 \left( c_2 \left(c_1 +c_s\right)m_1+\left(c_1+c_2\right)c_s
m_s\right)+ \right. \nonumber \\  & &  \left. 16 \pi^4 f^4 c_1 c_2 c_s
m_1 m_s\right) \left/ \right.\left(-1 + 4 \pi^2 f^2 \left(c_1 m_1 +
c_s m_1 + c_1 m_2 + c_2 m_2 + \right. \right. \nonumber \\ & & \left. \left.  
c_s m_2 + c_s m_s\right) + 16 \pi^4 f^4 \left(
c_1 c_2 m_1 m_2 - c_2 c_s m_1 m_2 - c_1 c_s m_1 m_s -  \right. \right. 
\nonumber \\ & & \left. \left. c_1 c_s m_2 m_s -
c_2 c_s m_2 m_s \right) + 64 \pi^6 f^6c_1 c_2 c_s m_1 m_2 m_s \right).
\end{eqnarray}
In the above, $c_1$ is the reciprocal spring constant $k_1$ for the
spring separating the sphere from the intermediate mass of the
transducer, $c_2$ is the reciprocal spring constant $k_2$ for the
spring separating the intermediate mass from the transducer mass, $c_s$
is the reciprocal spring constant $k_s$ of the effective spring
separating the effective mass of the sphere from mechanical ground,
$m_1$ is the mass of the intermediate mass of the transducer, $m_2$ is
the mass of the transducer mass, and $m_s$ is the effective mass of the
sphere for the lowest quadrupole mode.  The masses and springs in the
transducer are shown in Fig.~\ref{fig:sphere}.  The spring constants
can be written for each stage in the transducer (j=1,2, or $s$)
\begin{equation}
k_{j} = \left(2 \pi f_0\right)^2 m_j + i \left(2 \pi f\right)
\left(2 \pi f_0\right) m_j / Q_j,
\end{equation}
where $Q_j$ is the quality factor of the appropriate stage of the
transducer and sphere.  The $Q$'s depend on composition, temperature,
and the connections between the masses and springs.  The amplifier
force noise is shown in Fig.~\ref{fig:sphereparts} graphed as a strain
spectrum .

The thermal noise of the sphere can be written~\cite{Harry}
\begin{equation}
S_{\mathrm{sph,therm}} = 2 k_B T
\Re\left(y_{22}\left(f\right)\right),
\end{equation}
where $T$ is the physical temperature of the sphere.  We modeled the
antenna as having a $T$ of 50~mK, although 95~mK is the lowest a bar
has been cooled at equilibrium~\cite{Coccia,Astone}.  The term
$\Re\left(y_{22}\left(f\right)\right)$ depends on the $Q$'s, with
higher $Q$'s resulting in lower thermal noise.  The sphere and
intermediate mass were modeled in aluminum and the transducer mass was
modeled in niobium. Each of the mechanical $Q$'s was modeled as $40
\times 10^6$~\cite{Duffy90,Duffy94,Geng}. Depending on the design of
the transducer, the final $Q$ can be degraded by the addition of loss
from electrical coupling to the SQUID circuit~\cite{thesis}.  The
sphere's thermal noise is shown in Fig.~\ref{fig:sphereparts} graphed
as a strain spectrum.

All of the noise sources for the sphere were then combined to determine
the total noise;
\begin{equation}
S_{\mathrm{tot}}\left(f\right) = S_u \left(f\right) + S_{f,\,
\mathrm{out}}\left(f\right)+S_{\mathrm{sph,therm}}.
\label{eqn:sphtot}
\end{equation}
This total noise is shown, along with each component, in
Fig.~\ref{fig:sphereparts} graphed as a strain spectrum.  

The gravitational wave signal is applied as force on the
spherical antenna, but is read out as velocity of the transducer
mass.  Thus, before a signal can be compared to these noise
sources, the gravitational strain must be converted to a force
and be passed through the admittance matrix of the
sphere-transducer system.  The comparable signal can be
written~\cite{Harry}
\begin{equation}
\Sigma\left(f\right) = \frac{\pi Y^{5/2} \Pi m_s}{f_{0}^{3}
\rho^{3/2}} f^2 |y_{21}\left(f\right) h\left(f\right) |^2,
\label{eqn:comsig}
\end{equation}
where $Y$ is Young's modulus for the sphere material, $\rho$ is the
density of the sphere material, $\Pi$ is the reduced cross section of
the sphere and equals 0.215 for the lowest quadrupole mode~\cite{Zhou},
$h\left(f\right)$ is the frequency-domain amplitude of the
gravitational wave, and the admittance matrix element
$y_{21}\left(f\right)$ can be written
\begin{eqnarray}
y_{21}\left(f\right) & = & 2 \pi f c_s i \left/ \right. \left( -1
+ 4 \pi^2 f^2 \left(c_1 m_1 + c_s m_1 + c_1 m_2 + c_2 m_2 + c_s m_2 + c_s m_s
\right) \right. \nonumber \\ & & \left. - 16 \pi^4 f^4 \left( c_1 c_2 m_1
m_2 + c_2 c_s m_1 m_2 + c_1 c_s m_1 m_s
+ c_1 c_s m_2 m_s + c_2 c_s m_2 m_s \right)  \right. \nonumber \\ & & \left.
+ 64 \pi^6 f^6 c_1 c_2 c_s m_1 m_2 m_s \right).
 \end{eqnarray}
Table~\ref{table:sphere} shows all the parameters used in the
sphere model.

\begin{figure}
\caption{Schematic drawing of an interferometric gravitational
wave detector equipped for resonant sideband extraction.  The laser
creates a light beam which is sent through the power recycling mirror
into two arms of a Michelson interferometer.  The signal exits at the
output port through the signal recycling mirror, which forms an
additional Fabry-Perot cavity with the Michelson interferometer.
Finally, the signal is detected past the signal recycling mirror with a
photodiode.}
\begin{center}
\epsfxsize=16cm \leavevmode \epsfbox{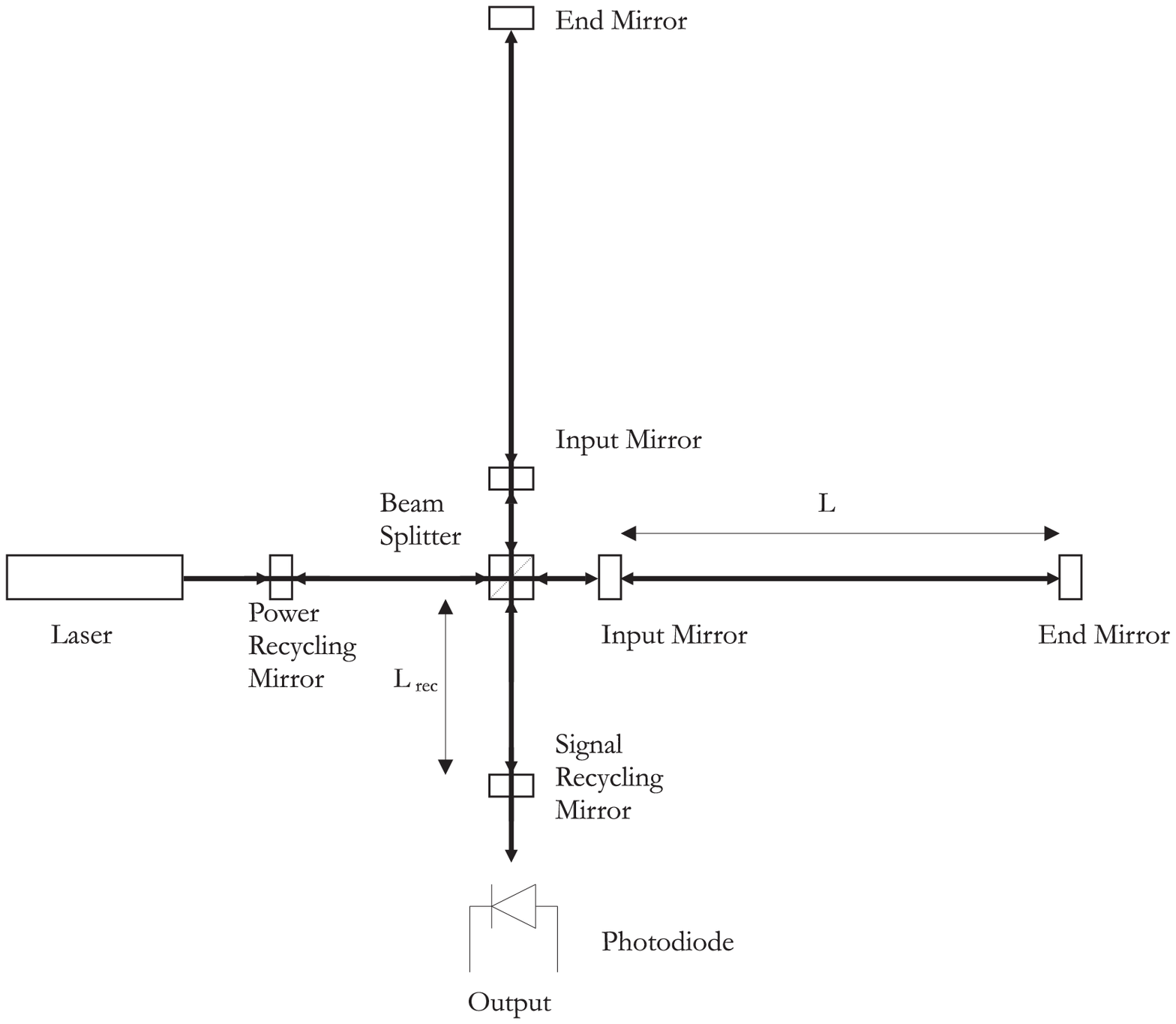}
\end{center}
\label{fig:ligo}
\end{figure}

The interferometer was modeled using a slightly modified version of the
{\tt BENCH} program~\cite{bench}.  The noise in the interferometer is
dominated by three types of noise; seismic, thermal, and optical
readout noise.  Each noise source was modeled using parameters from the
advanced LIGO white paper~\cite{ligoII}, which is scheduled to be
implemented in 2005. The corresponding advanced interferometer is
scheduled to begin taking data in 2007.  A schematic drawing of LIGO
with RSE is shown in Fig.~\ref{fig:ligo}

Seismic noise is expected to dominate the advanced LIGO noise budget at
low frequencies.  To reduce the effect of seismic noise, each element
of the interferometer will be supported by a four-stage suspension
which in turn is supported from a vibration isolation stack. This
vibration isolation will consist of two stages of six-degree of freedom
isolation. It will use a combination of active and passive isolation
with an external hydraulic actuation stage. The isolation was designed
to make seismic noise negligible compared to other noise sources above
some $f_{\mathrm{seismic}}$, expected to be 10~Hz.  In the model,
seismic noise was made extremely high below 10~Hz and vanishingly small
above this frequency.

Thermal noise will be the dominant noise source in advanced LIGO in the
intermediate frequency band above 10~Hz.  This noise can be divided
into two types; thermal noise from the internal degrees of freedom of
the interferometer mirrors, and thermal noise from the suspension that
supports the mirrors. The mirrors are planned to be made of m-axis
sapphire, 28~cm in diameter and 30~kg in mass. Sapphire has been found
to have much lower internal friction than fused
silica~\cite{Braginsky,Rowan}, which is used in initial LIGO. However,
sapphire suffers from much higher thermoelastic damping than
silica~\cite{Braginsky2000}.

The internal mode thermal noise from the sapphire mirror comes from
structural damping and thermoelastic damping.  The noise from
structural damping can be found from the loss angle $\phi$ by
\begin{equation}
S_{\mathrm{str}}\left(f\right) = \frac{1}{L^2} \frac{8 k_{B} T
\phi}{\pi f} \left({\mathcal C}_1 + {\mathcal C}_2\right),
\end{equation}
where $k_{B}$ is Boltzmann's constant, $T$ is the temperature, $f$ is
the frequency, and $L$ is the interferometer arm length. The constants
${\mathcal C}_1$ and ${\mathcal C}_2$ are the overlap between the
normal modes of the mirrors and the gaussian-profile laser, which has a
width $w_1$ at the input mirror and $w_2$ at the end mirror. They are
found from~\cite{bondu,liu}
\begin{eqnarray}
{\mathcal C}_j = \frac{\left(1-\sigma^2\right)}{\pi r Y} & &
\sum_{i=1}^{N} \left(\frac{\exp\left(-\left(\zeta_i
w_j/(2r)\right)^2\right)} {\zeta_i {\mathrm
J}_{0}\left(\zeta_i\right)^2}\frac{\left(1- \exp\left(-4 \zeta_i
h/r\right) +4 \zeta_i h/r \exp\left(-2 \zeta_i h/r\right) \right)}
{\left(1-\exp\left(-2 \zeta_i h/r\right)\right)^2 - 4
\left(\zeta_i
h/r\right)^2 \exp\left(-2 \zeta_i h/r\right)}\right) \\
\nonumber & + & \frac{r^2}{6 \pi h^3 Y} \left( h^4/r^4 + 12 \pi
h^2 \sigma \sum_{i=1}^{N} \frac{\exp\left(-\left(\zeta_i
w_j/(2r)\right)^2/2\right)} {\zeta_i^2 {\mathrm
J}_{0}\left(\zeta_i\right)} \right. \\ \nonumber & + & \left. 72
\left(1-\sigma\right) \left( \sum_{i=1}^{N}
\frac{\exp\left(-\left(\zeta_i w_j/(2r)\right)^2/2\right)}
{\zeta_i^2 {\mathrm J}_{0}\left(\zeta_i\right)}\right)^2\right).
\end{eqnarray}
where $\sigma$ is the Poisson ratio of the mirror material, $Y$ is it
Young's modulus, $r$ is the radius of the mirror, $\ell$ is the
thickness of the mirror, $w$ is the Gaussian beam width of the laser at
the mirror, $\zeta_i$ are the zeros of the first order Bessel function
J$_{1}$, and J$_{0}$ is the zeroth order Bessel function.  The value of
$\phi$ used is the lowest value measured for a piece of
sapphire~\cite{Braginsky}. The thermal noise effects of making a
sapphire piece into a mirror are under study, but the polishing and
especially coating of the mirror are expected to cause some excess
loss~\cite{coating,sheilaaspen}.  

Thermoelastic damping also contributes to thermal noise from the
mirrors.  It is found, in the limit of large mirror diameter,
from~\cite{Braginsky2000}
\begin{equation}
S_{th}\left(f\right) = \left(\frac{\left(1 + \sigma\right)
 \, \alpha  \, T}{\pi \, f \, L \, C_V
\, \rho}\right)^2 \frac{16 \, \kappa \, k_B}{\sqrt{\pi}} \left(
1/w_1^3 +1/w_2^3\right),
\end{equation}
where $\alpha$ is the thermal expansion coefficient, $C_V$ is the
heat capacity at constant volume, $\rho$ is the density, and
$\kappa$ is the thermal conductivity.  Fused silica is available as a back
up material which does not have as much thermoelastic loss and has
recently been shown to have a $\phi$ as low as $1.8 \times 10^{-8}$ in
certain circumstances~\cite{highq}.

Thermal noise from the suspension, which supports the mirrors below the
vibration isolation stack, will be reduced in advanced LIGO by
replacing the steel slings with fused silica ribbons.  Fused silica has
much less internal friction than
steel~\cite{Kovalik,Gretarsson,Startin}, although with ribbon geometry
surface loss limits the achievable
dissipation~\cite{Gretarsson,ribbon}. Thermal noise from a ribbon
suspension with surface loss has recently been considered~\cite{ribbon}
and the results give thermal noise, expressed as gravitational wave
stress squared per Hertz, as
\begin{equation}
S_{\mathrm{ susp}} \left(f\right) = 32 k_B T \phi_{\mathrm{dil}}
\, g\left/L^2 \left(L_{\mathrm{sus}} \, m \, 2 \pi f \left(
\left(\left(2 \pi f\right)^2 - \omega_{\mathrm{pen}}^2 \right)^2
+ \omega_{\mathrm{pen}}^4 \phi_{\mathrm{dil}}^2\right) \right)\right.,
\end{equation}
where $g$ is the acceleration due to gravity, $L_{\mathrm{sus}}$
is the length of the suspension, $m$ is the mass of the mirror,
$\omega_{\mathrm{pen}}$ is the angular frequency of the pendulum
mode, and $\phi_{\mathrm{dil}}$ is the diluted loss angle. This
diluted loss angle is defined as
\begin{equation}
\phi_{\mathrm{dil}} = \sqrt{Y \left/\left(12 \gamma \sigma
L_{\mathrm{sus}}^2\right) \right.} \; d \left( \phi_{\mathrm{th}}
+ \phi_{\mathrm{int}} \right),
\end{equation}
where $Y$ is Young's modulus for the ribbon material, $\gamma$ is
the ratio of stress in the ribbon to its breaking stress,
$\sigma$ is Poisson's ratio for the ribbon material, $d$ is the
ribbon thickness, $\phi_{\mathrm{th}}$ is the loss angle due to
thermoelastic damping, and $\phi_{\mathrm{int}}$ is the loss
angle due to internal friction in the ribbon.  Thermoelastic
damping in ribbons is found from~\cite{Zener}
\begin{equation}
\phi_{\mathrm{th}} = \frac{Y \alpha^2 T}{C} \frac{2 \pi f
\tau_d}{1+\left(2 \pi f \right)^2 \tau_d^2},
\end{equation}
where $\alpha$ is the thermal expansion coefficient of the ribbon
material, $C$ is the heat capacity per unit volume, and $\tau_d$
is the time constant for thermal diffusion which in ribbons is
given by
\begin{equation}
\tau_d = d^2/\left(\pi^2 D\right),
\end{equation}
with $D$ being the thermal diffusion coefficient for the ribbon
material.  The internal friction in a thin ribbon is given
by~\cite{Gretarsson}
\begin{equation}
\phi_{\mathrm{int}} = \phi_{\mathrm{bulk}} \left( 1 + 6 d_s/d
\right),
\end{equation}
where $\phi_{\mathrm{bulk}}$ is the loss angle in the bulk of the
ribbon material, and $d_s$ is the dissipation depth that characterizes
the excess loss arising from the surface of the ribbon.  The numbers
for $\phi_{\mathrm{bulk}}$ and $d_s$ in Table~\ref{table:ligo}
represent possibly achievable values, lower values for both have been
observed~\cite{highq,glasgowribbon}.  Determining realizable values for
these parameters in advanced LIGO is an area of intense research.

Optical readout noise in the interferometer can be evident at any
frequency in the LIGO detection band.  This noise source has two
separate components: radiation pressure noise from the pressure
exerted on the mirrors by the laser and shot noise from the
inherent granularity (photons) of the laser light. These two
noise sources are complementary to each other, both depend on the
laser power.  Recently, optical readout noise in a signal-recycled 
interferometer has been considered from a fully quantum mechanical
perspective~\cite{Buonanno}.  The noise spectrum does differ from 
the one we calculate here, but the difference at high frequencies in 
a narrowband configuration are negligible.

The optical power stored in the interferometer is an important
parameter for the optical readout noise.  There are a number of
optical cavities in LIGO formed by the different mirrors (input
mirrors, end mirrors, power recycling mirror, signal recycling
mirror, etc.) and each one stores a different amount of power. It
is convenient to quote a single power, the power incident on the
beam splitter, and then calculate the power in different cavities
in terms of this single value.  The power at the beam splitter is
proportional to the power out of the laser, $P$, through the
power recycling factor
\begin{equation}
P_{bs} = G_{pr} P,
\end{equation}
which is found from
\begin{equation}
G_{pr} = 1/\left(2 N \beta + \alpha_{\mathrm{BS}} \right),
\end{equation}
where $\alpha_{\mathrm{BS}}$ is the fractional power loss at the
beam splitter, $\beta$ is the fractional power loss at each
mirror, and $N$ is the number of bounces that the light makes in
each arm, on average.  In Fabry-Perot cavities, in the large
finesse limit, the value $N$ can be found from the finesse,
\begin{equation}
N = 2 \mathcal{F}/\pi,
\end{equation}
where the finesse $\mathcal{F}$ is found from the amplitude
transmittance of the input mirror, $t_1$:
\begin{equation}
{\mathcal F} = 2 \pi/(t_1^2 + 2 \beta).
\end{equation}
These equations taken together with the parameters in
Table~\ref{table:ligo} give the power at the beam splitter,
\begin{equation}
P_{bs} = 9.3~\mathrm{kW}. \label{eqn:pbs}
\end{equation}
This power must be kept from being too high because absorption of light
in the transmitting mirrors, beam splitter, and coatings can lead
to thermal lensing.  The acceptable thermal lensing limit can be
calculated from (including a factor of 2 safety margin)
~\cite{thermallens}
\begin{equation}
P_{\mathrm{max}} \approx \frac{\kappa}{2 \,
\mathrm{d}n/\mathrm{d}T} \, \frac{\lambda}{1.43 A_g
t_{\mathrm{BS}} + 1.3 A_g \ell + \frac{1}{2} N a_{\mathrm{coat}}},
\end{equation}
where $\kappa$ is the thermal conductivity of the substrate material,
$\mathrm{d}n/\mathrm{d}T$ is the change in index of refraction of the
substrate with temperature, $A_g$ is the optical absorption of the
substrate, and $a_{\mathrm{coat}}$ is the relative absorption of the
optical coating.  Using the numbers in Table~\ref{table:ligo}, this
maximum allowable power is
\begin{equation}
P_{\mathrm{max}} = 750~\mathrm{W}.
\end{equation}
In order to realize the higher power in Eq.~(\ref{eqn:pbs}), a
correction scheme must be utilized that increases $P_{\mathrm{max}}$ by
a factor of at least 12.4 for sapphire optics. Research is underway to
have such a correction scheme available for advanced LIGO~\cite{ligoII}.

Both parts of the optical readout noise depend on the response of
the coupled cavity system in the interferometer. This response
can be described by the transfer function between the amplitude
of the light in the arm cavity and the amplitude of light that
enters through the input mirror in each sideband~\cite{Buonanno}:
\begin{eqnarray}
G_{0,1} &=& 1 - r_1 r_2 \exp\left(i \tau_a 2 \pi f \right) - r_1
r_3 \exp\left(i \left(\tau_s 2 \pi f +  \delta\right)\right) +
r_2 r_3 \exp\left(i\left( \left(\tau_a+\tau_s\right) 2 \pi f
+\delta\right)\right), \label{g01} \\
G_{0,2} &=& 1 - r_1 r_2 \exp\left(-i \tau_a 2 \pi f \right) - r_1
r_3 \exp\left(-i \left(\tau_s 2 \pi f + \delta\right)\right) +
r_2 r_3 \exp\left(-i\left( \left(\tau_a+\tau_s\right) 2 \pi f
+\delta\right)\right) \label{g02}
\end{eqnarray}
where $r_1$,$r_2$, and $r_3$ are the amplitude reflection
coefficients at the input mirror, the end mirror, and the signal
recycling mirror, respectively, $\tau_a (=2L/c)$ is the light
transit time between the input mirror and the end mirror,
$\tau_s(=2 L_{\mathrm{rec}}/c)$ is the light transit time between
the input mirror and the signal recycling mirror with
$L_{\mathrm{rec}}$ the length of this signal recycling cavity,
and $\delta$ is the phase accumulated by the reflected light
coming off the signal recycling mirror due to its position. The
amplitude reflection coefficient for the input mirror can be
found from
\begin{equation}
r_{1}^{2} = 1 - t_{1}^{2} - \beta.
\end{equation}
The amplitude reflection coefficient for the end mirror can be
found from
\begin{equation}
r_{1}^{2} = 1  - \beta.
\end{equation}
The amplitude reflection coefficient for the signal recycling mirror,
$r_{3}$, is a tunable parameter as is the accumulated phase, $\delta$.

Radiation pressure noise is largest at low frequencies and, for
initial LIGO,
is masked by other low frequency noise (seismic and suspension thermal
noise)~\cite{Abramovici}.  In advanced LIGO, the suspension thermal noise may be low enough that radiation pressure is important, but it will still
not be the dominant noise source. Radiation pressure was modeled by
\begin{equation}
S_{\mathrm{rad}}\left(f\right) = 32 P_{bs} \,  2 \pi f_\lambda \,
\hbar \  t_{1}^4 \, t_3^2 \, r_2^2 \left(
1/|G_{0,1}|+1/|G_{0,2}|\right)^2\left/\left(\left( 1- r_1
r_2\right) (2 \pi f)^2 c \, m \, L \,  \right)^2\right.,
\end{equation}
where $f_\lambda$ is the frequency of laser light and $t_{3}$ is
the amplitude transmittance of the signal recycling mirror.

Shot noise from the laser is the dominant noise source at high frequencies
for intial LIGO, and this will continue for advanced LIGO.  This noise
source was modeled by
\begin{equation}
S_{\mathrm{shot}}\left(f\right) = \left(\frac{f \left(1-r_1
r_2\right)}{f_\lambda \sin\left(\pi f \tau_a\right) t_1^2 r_2 t_3
\left(1/|G_{0,1}|+1/|G_{0,2}|\right) }\right)^2 \frac{4 \pi f
\hbar}{\eta P_{bs}},
\end{equation}
where $\eta$ is the quantum efficiency of the photodiode.

\begin{figure}
\caption{Strain spectra for narrowband interferometers and
spherical, resonant mass antennas. \\(A) Both interferometer and sphere
are maximally sensitive at 795~Hz, corresponding to the quadrupole mode
of an aluminum sphere with diameter 3.25~m and a phase shift $\delta =
0.2271$ for the interferometer. \\(B) Strain spectra for four
narrowband interferometers, sensitive at 795~Hz, 1100~Hz, 1520~Hz, and
2067~Hz. Also shown are the strain spectra for the four spherical,
resonant mass antennas with the same resonance frequencies. The spheres
are less sensitive than the interferometers at the resonance point, but
have roughly the same sensitivity as the interferometers
off-resonance.}
\hspace{-1.5em}(A)
\begin{center}
\epsfxsize=11cm \leavevmode \epsfbox{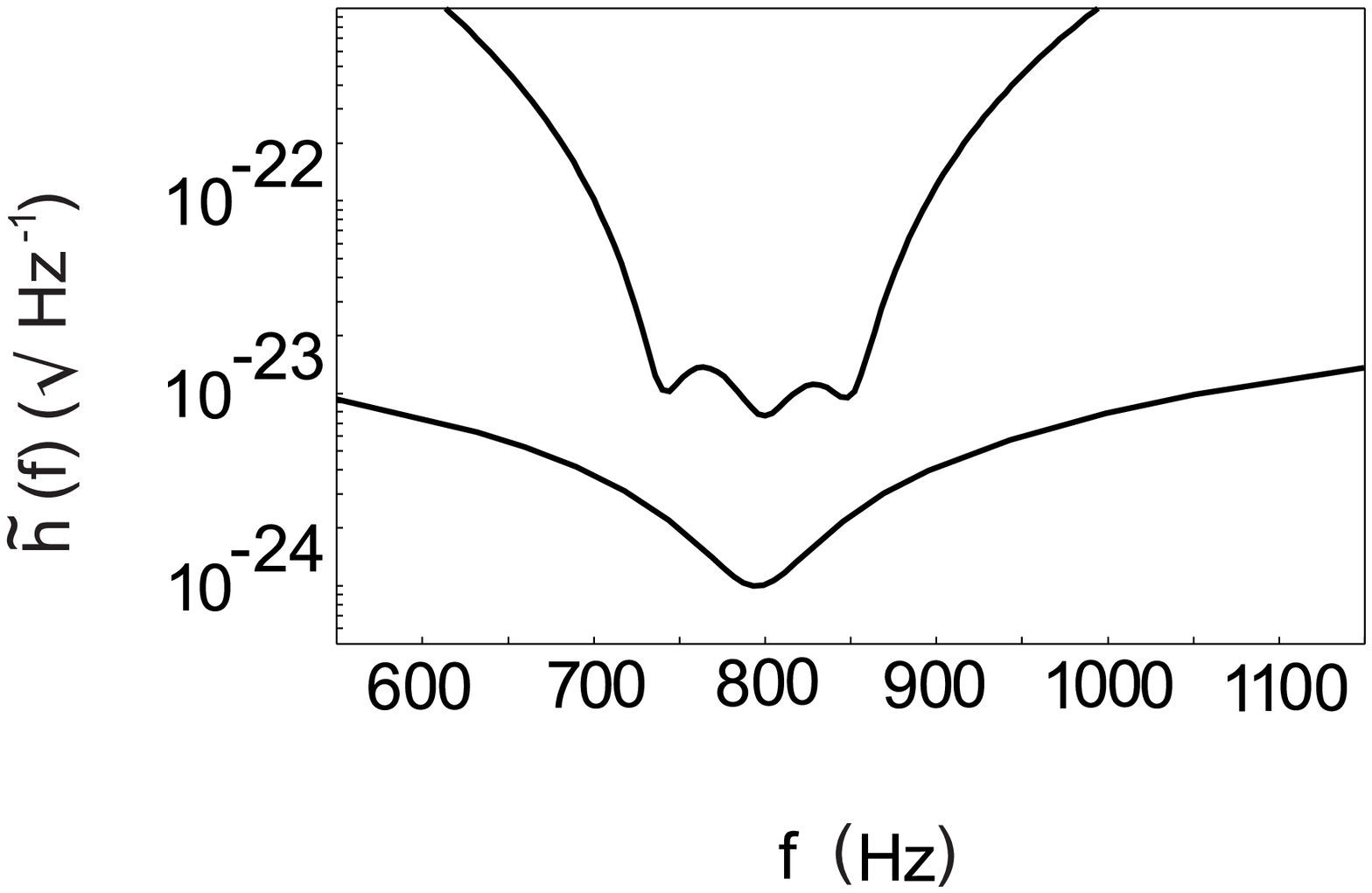}
\end{center}
(B)
\begin{center}
\epsfxsize=11cm \leavevmode \epsfbox{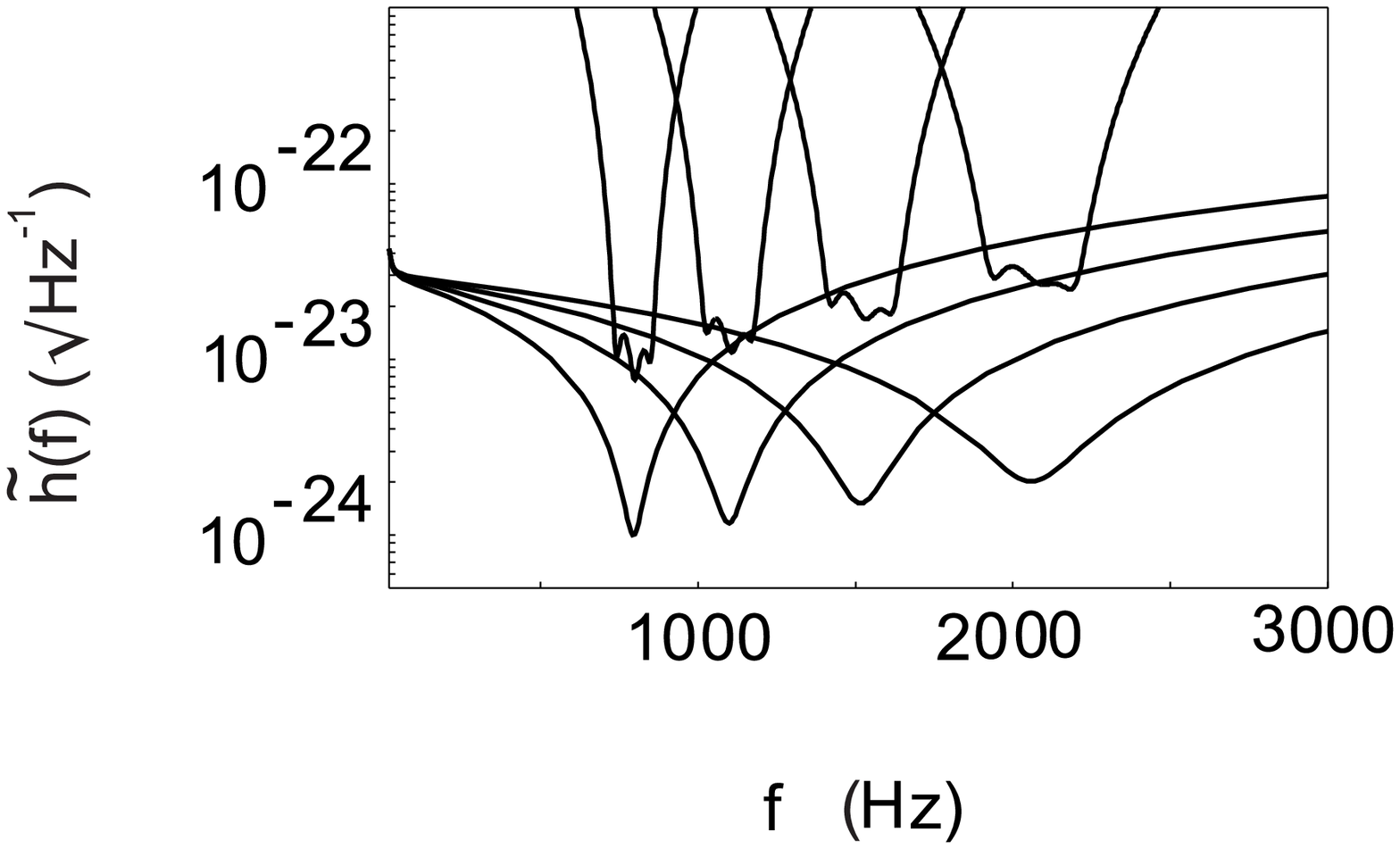}
\end{center}
\label{fig:strain}
\end{figure}

All of these noise sources were combined to create the total noise curve for advanced LIGO;
\begin{equation}
S_{\mathrm{tot}}(h) =  S_{\mathrm{seis}} + S_{\mathrm{int}}+
S_{\mathrm{susp}} + S_{\mathrm{shot}} + S_{\mathrm{rad}}.
\label{eqn:totnoise}
\end{equation}
To produce a numerical estimate of the noise and then a signal-to-noise
ratio for a given source, values must be provided for all the
parameters that go into Eq.~(\ref{eqn:totnoise}).  We used values from
the advanced LIGO white paper~\cite{ligoII} as much as possible.  The
values chosen for all relevant parameters are shown in
Table~\ref{table:ligo}. A graph of advanced LIGO's noise compared with
spheres is shown in Fig.~\ref{fig:strain}.

The masses in LIGO are designed to be as close to being in local free
fall in the sensitive direction as possible. Therefore, the strain from
a passing gravitational wave directly gives the change in position of
the mirrors.  The comparable signal, similar to Eq.~(\ref{eqn:comsig}),
for an interferometer reads
\begin{equation}
\Sigma\left(f\right) = |h\left(f\right)|^2. \label{eqn:intsource}
\end{equation}

Using Eqs.~\ref{eqn:sphtot} and~\ref{eqn:totnoise} for the noise of a
sphere and interferometer and Eqs.~\ref{eqn:comsig}
and~\ref{eqn:intsource} as the comparable signals, the signal-to-noise
ratio density for each detector can be found from
\begin{equation}
\sigma \left(f\right) = \Sigma\left(f\right) /
S_{\mathrm{tot}}\left(f\right).
\end{equation}
Integrating the signal-to-noise ratio density gives the signal-to-noise
ratio;
\begin{equation}
S/N =\left\langle \int_{-\infty}^{+\infty}
\sigma\left(f\right)
{\mathrm d}f \right\rangle, \label{eqn:snr}
\end{equation}
where the angle brackets, $\langle \ldots \rangle$, denote averaging
over gravitational wave polarization and direction.  This results in a
factor of $1/5$~\cite{Thorne87} for interferometers and a factor of 1
for spheres, as spheres are always optimally oriented. This value,
$S/N$, is the figure of merit for a gravitational wave detector and
will be used to compare the effectiveness of these two different
approaches to the different astronomical sources.

\begin{figure}
\caption{Strain spectra for an interferometeric gravitational wave detector
with resonant sideband extraction showing all the components of the
noise. The dash-dotted-dotted line shows the shot noise from a 125~W
laser, the dashed line shows the sapphire mirrors' internal mode thermal
noise, and the dashed-dotted line shows the thermal noise from the ribbon
suspension, and the dotted line shows the radiation
pressure. The solid line is the total noise. \\(A)The noise components
when the interferometer is in a narrowband configuration tuned to
795~Hz. \\(B)The noise components when the interferometer is in a
broadband configuration optimized for binary neutron star inspiral.}
(A)
\begin{center}
\epsfxsize=15cm \leavevmode \epsfbox{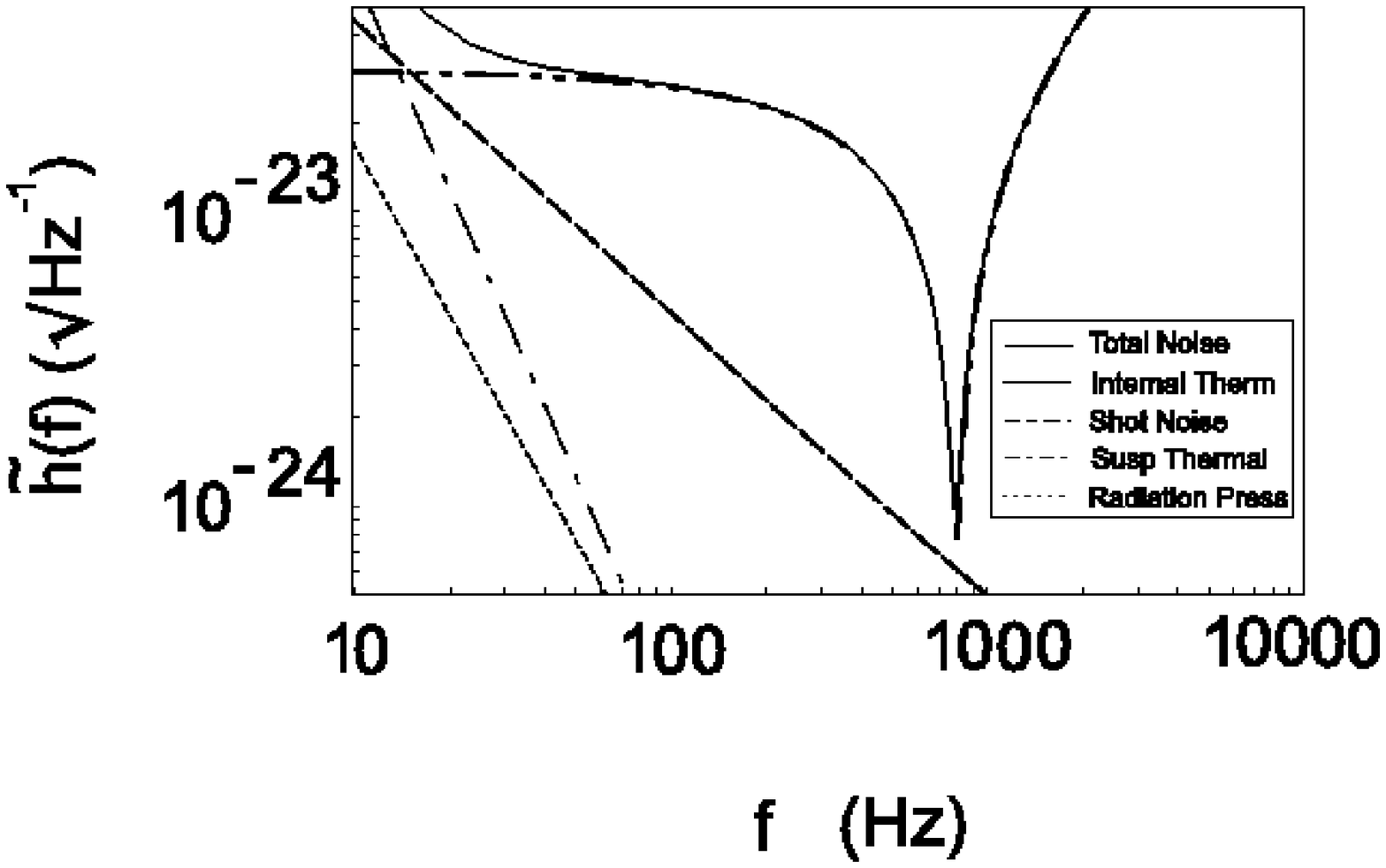}
\end{center}
(B)
\begin{center}
\epsfxsize=15cm \leavevmode \epsfbox{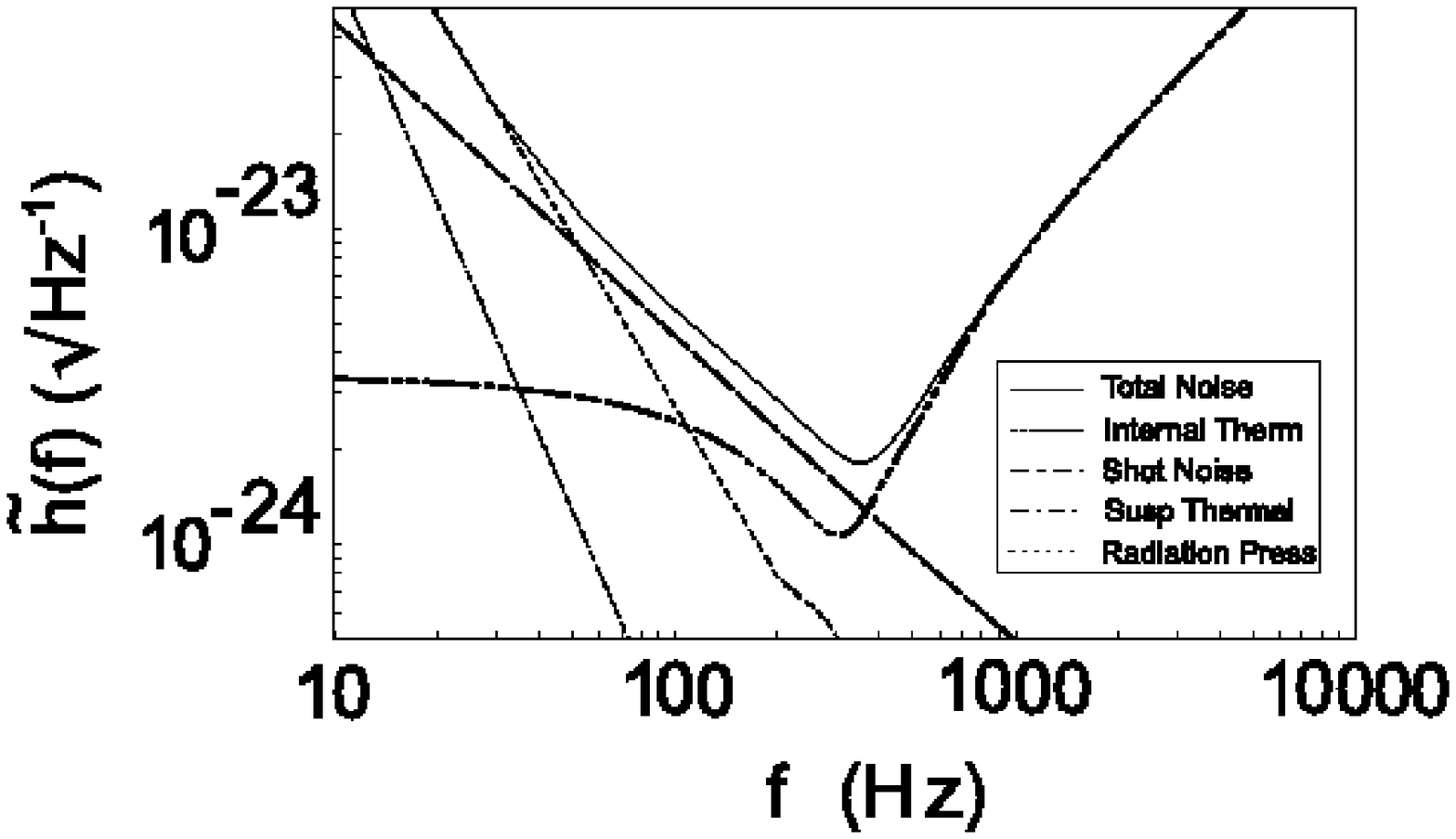}
\end{center}
\label{fig:ligoparts}
\end{figure}

To compare the sensitivities of the the antennas, it is useful to
calculate the strain spectral density $\tilde{h}\left(f\right)$,
\begin{equation}
\tilde{h}\left(f\right) = \sigma \left(f\right)/|h\left(f\right)|^2.
\end{equation}
This quantity encorporates both the total noise of the detector and
the cross section.  It is the strain spectrum that is shown in
Figures~\ref{fig:sphereparts},\ref{fig:strain} and~\ref{fig:ligoparts}.

\section{Sources}

One category of sources for gravitational radiation at high frequencies
(above 700~Hz) is from internal motion of compact bodies such as
neutron stars.  The coalescence and merger of neutron stars as well as
neutron star formation in collapsing stellar cores are promising
sources of detectable gravitational waves.  Detecting and analyzing
these waves promises to teach us a great deal about the physics of
strong gravitational fields and extreme states of
matter~\cite{Thorne95}. Because of the high rotational velocities and
strong gravitational fields present in such compact objects, numerical
simulations must include the effects of general relativity to model the
system dynamics realistically enough for use in analysis of the data
from antennas. To accomplish this goal, a three-dimensional Smoothed
Particle Hydrodynamics code~\cite{HK89} has been modified to include
the General Relativistic ``back reaction''~\cite{Houser00a,Houser2000}.
The gravitational radiation from these systems is calculated in the
quadrupole approximation.

The component stars of a widely separated binary neutron star
system will spiral together due to orbital energy losses by
gravitational radiation reaction, and eventually
coalesce~\cite{P91,NPS91,FC93}.  Because neutron stars have
intense self-gravity, as they inspiral they do not
gravitationally deform one another significantly until several
orbits before final coalescence \cite{Thorne95}.  When the binary
separation is comparable to a neutron star radius, tidal
distortions become significant, hydrodynamical effects become
dominant, and coalescence occurs in a few orbits.

The inspiral phase of the evolution comprises the last several thousand
binary orbits and covers the frequency range $f \sim 10$ -- $1000
\;{\rm Hz}$.  The final coalescence is believed to emit its
gravitational waves in the kilohertz frequency band range $800\; {\rm
Hz} < f < 2500\; {\rm Hz}$~\cite{Thorne95,Cutler93,CF94,Thorne96a}. The
observation of the inspiral and coalescence waveforms will reveal
information about the masses and spin angular momenta of the bodies,
the initial orbital elements of the system, the neutron star radii and
hence the equation of state for nuclear
matter~\cite{Thorne95,Cutler93,CF94,Thorne96a}.

Theoretical estimates of the formation rates for binary neutron star
systems---with tight enough orbits to merge due to gravitational
radiation within a Hubble time---can be obtained from empirical rate
estimates based on the observed sample~\cite{Kalogera99a}.  The most
recent study gives a Galactic lower and upper limit of $2 \times
10^{-7}\; {\rm yr}^{-1}$ and $\sim 6-10 \times 10^{-6}\; {\rm
yr}^{-1}$, respectively \cite{KalogeraNarayan}. Alternatively, by
modeling the evolution of the Galaxy's binary star population, the best
estimates for coalescence events have been estimated to be as high as $3
\times 10^{-4}$ coalescences per year in our galaxy, and several per
year out to a distance of $60\; {\rm Mpc}$ \cite{Thorne95}.  We have
used 15~Mpc as an optimistic estimate and 200~Mpc as a pessimistic
estimate of the distance antennas will need to look to get about one
event per year for this source.

For the simulation presented here, equal mass component stars are used.
Each star is assumed to have a total mass of $M_t = 1.4\; M_\odot$, and
equatorial radius, $R_{\rm eq} = 10\; {\rm km}$, where $M_\odot$ is one
solar mass.  The star is modeled as a differentially rotating
axisymmetric fluid which use a polytropic equation of state,
\begin{eqnarray}
P & = & k \rho^{\gamma} \\
  & = & k \rho^{1+1/n},
\end{eqnarray}
where $k$ is a constant that measures the specific entropy of the
material and $n$ is the polytropic index.  The value $n=1$ is used to
simulate cold nuclear neutron star matter. Each star rotates
counterclockwise about the $z-$axis with an equatorial surface speed of
approximately $0.083 \; c$~\cite{Houser2000}.  

Since the binary system spirals together due to energy and
angular momentum losses via the emission of gravitational
radiation, as the orbit decays, it circularizes radially.  Thus,
it is a good approximation to assume a circular orbit provided
the system is expected to have existed for some
time~\cite{Schutz90}.  The component stars used in the simulation
are initially placed on a sufficiently wide circular orbit
(center of mass distance is $40\;{\rm km}$) so that tidal effects
are negligible~\cite{Zhuge94,Zhuge96}.  Because of the large
initial separation, the stars start out in the point mass regime,
and as a result, their waveforms can be compared directly to the
theoretical point-mass inspiral for two neutron stars~\cite{MTW}.

Although spin-orbit misalignment in coalescing compact binaries can
change the amplitude and modulation of the gravitational radiation
waveforms, this effect is believed unimportant in the case of a binary
neutron star system~\cite{Kalogera2000c}.  Thus, in the numerical
simulation, the orbital and spin angular momentum vectors were assumed
to be aligned.  If the magnetic axis is not aligned with the rotation 
axis, the ejecta from the coalescence can be trapped within the
field~\cite{BG96}.  The evolution of the magnetic field
configuration during the final inspiral phase of neutron star
binaries may have significant effects on the frequency and tidal
distortion of the coalescence, and hence on the gravitational
waveforms~\cite{AC96}.  The inclusion of this effect is the
subject of current research~\cite{Houser01}.  

\begin{figure}
\caption{Inspiraling and coalescing binary neutron stars. \\(A)
Particle positions for each neutron star during the coalescence
phase. The stars first fall together, reducing the gravitational wave
amplitude at twice the instantaneous orbital velocity. Then a bar forms
which creates an increased amplitude at twice the rotational velocity. 
\\(B)The frequency domain gravitational waveform averaged
over source orientation.  Notice the slight dip just above 1000~Hz
from the in-fall and the peak near 3500~Hz from the bar.  Each neutron
star was modeled as having a mass of $1.4~M_{\odot}$, an equatorial radius
of $10$~km, and a distance from the antenna of 15~Mpc.
}
(A)
\begin{center}
\epsfxsize=12cm \leavevmode \epsfbox{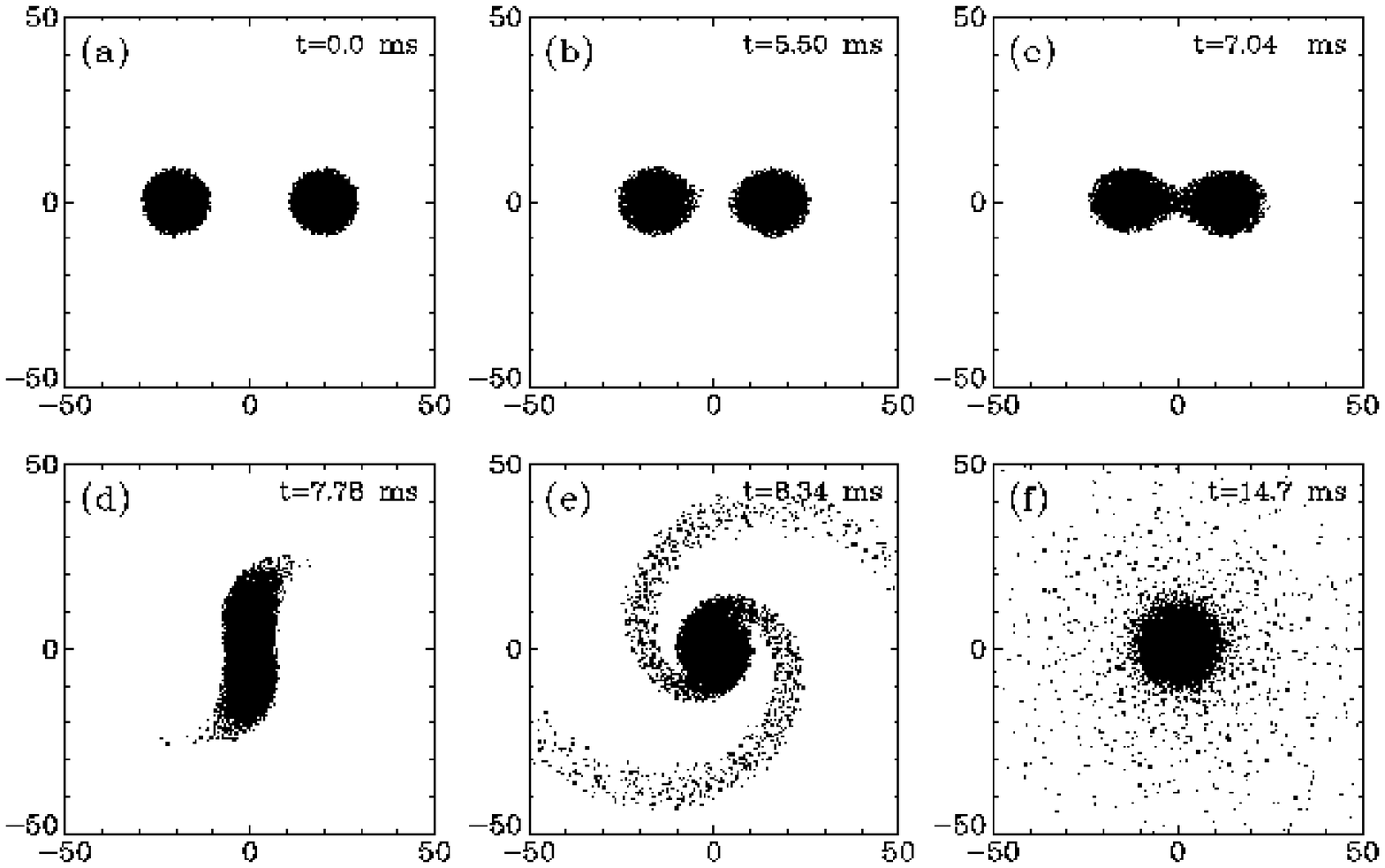}
\end{center}
(B)
\begin{center}
\epsfxsize=15cm \leavevmode \epsfbox{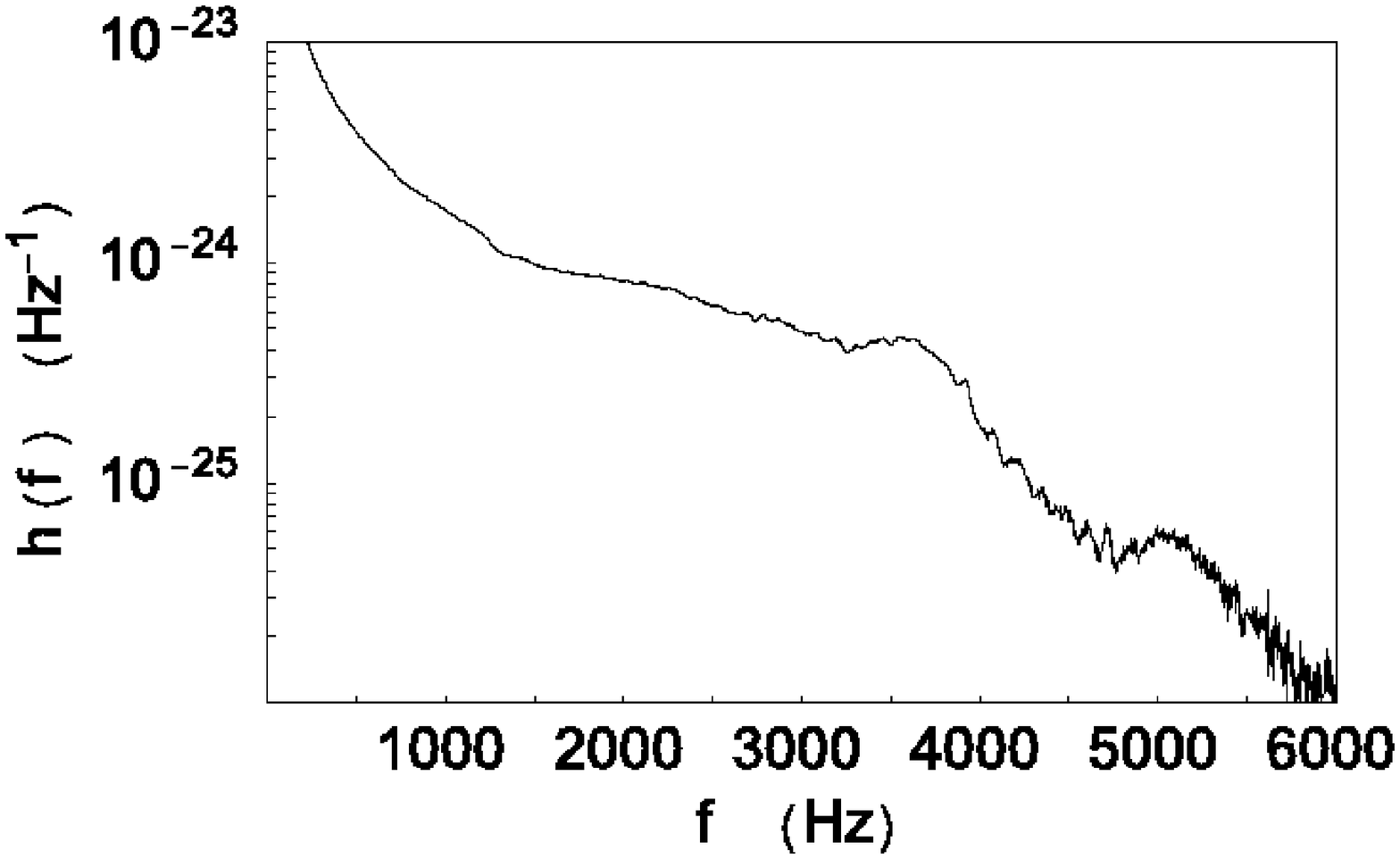}
\end{center}
\label{fig:sourcebns}
\end{figure}

A series of snapshots of the inspiral and coalescence of the 
neutron stars along with a graph of  the waveform generated is 
shown in Figure~\ref{fig:sourcebns}.  The waveform differs noticeably
from ones generated with purely Newtonian gravity~\cite{Zhuge94}.
The gravitational wave peak due to the bar formed during coalescence,
seen in Fig.~\ref{fig:sourcebns}(A)(d), is at a much higher frequency;  
3700~Hz with General Relativistic back reaction compared to 2500~Hz
in the Newtonian case.  This peak is also broader and less pronounced 
in the General Relativistic simulation than in the Newtonian.

Rotational instability during the gravitational collapse of a
massive star's degenerate core has long been considered an
interesting possible source of gravitational radiation.  A
typical scenario in which such a mechanism can operate is the
collapse of a rapidly rotating stellar core that has exhausted
its nuclear fuel and is prevented from collapsing to neutron star
size by centrifugal forces.  If a significant amount of angular momentum
remains in an initially axisymmetric core, collapse may be slowed or
temporarily stalled by centrifugal forces associated with
rotation.  If the core's rotation is large enough to strongly
flatten the core before, or as it reaches neutron-star density,
then an instability is likely to break the core's axial symmetry
\cite{Thorne95,DL77,Schutz86}.  The growth of such global
rotational instabilities provides a means for transport of
angular momentum out of the core into the surrounding envelope by
transforming the core into a bar-like configuration rotating
about the short axis, shedding mass in a spiral pattern, thereby
allowing collapse to continue to a supernova
\cite{Houser2000,Houser00a,Schutz86,LS95,SHC96,HC96,Houser1998}.

The strength of the gravitational waves from such a supernova
depends crucially on degree of non-sphericity during the
collapse, and somewhat on the speed of collapse.  If a
substantial fraction of the collapsing object's angular momentum
goes into generating gravitational rather than hydrodynamical
waves then the gravitational waves generated may be nearly as
strong as those generated from a coalescing binary
\cite{Thorne87}.

The event rates of Type II Supernovae are fairly well known from
observations \cite{Thorne87}.  In our Galaxy, Type II Supernovae
occur approximately once every 40 years, and several per year out
to the distance of the Virgo Cluster of Galaxies (about $10\;{\rm
Mpc}$).  Beyond this point, the rate increases roughly as the
cube of the distance, where by $\sim 300 {\rm \; Mpc}$ the
supernova rate becomes $\sim 10^4$ per year \cite{Thorne87,LS95}.
Although it is unclear what fraction of collapsing cores may
undergo an instability, if only $\sim 1/1000$ or $1/10^4$ do so,
this phenomena could be a significant source of detectable
gravitational radiation \cite{Thorne95}.

\begin{figure}
\caption{A rapidly rotating stellar core undergoing a dynamical
instability. \\(A)
Particle positions for the neutron star during the gravitational wave
emission which shows the bar shape that develops from the instability.
\\(B)The frequency domain gravitational waveform averaged
over source orientation.  The star was modeled as having a mass of
$1.4~M_{\odot}$, an equatorial radius of $10$~km, and a distance from
the antenna of 1~Mpc. }
(A)
\begin{center}
\epsfxsize=13cm \leavevmode \epsfbox{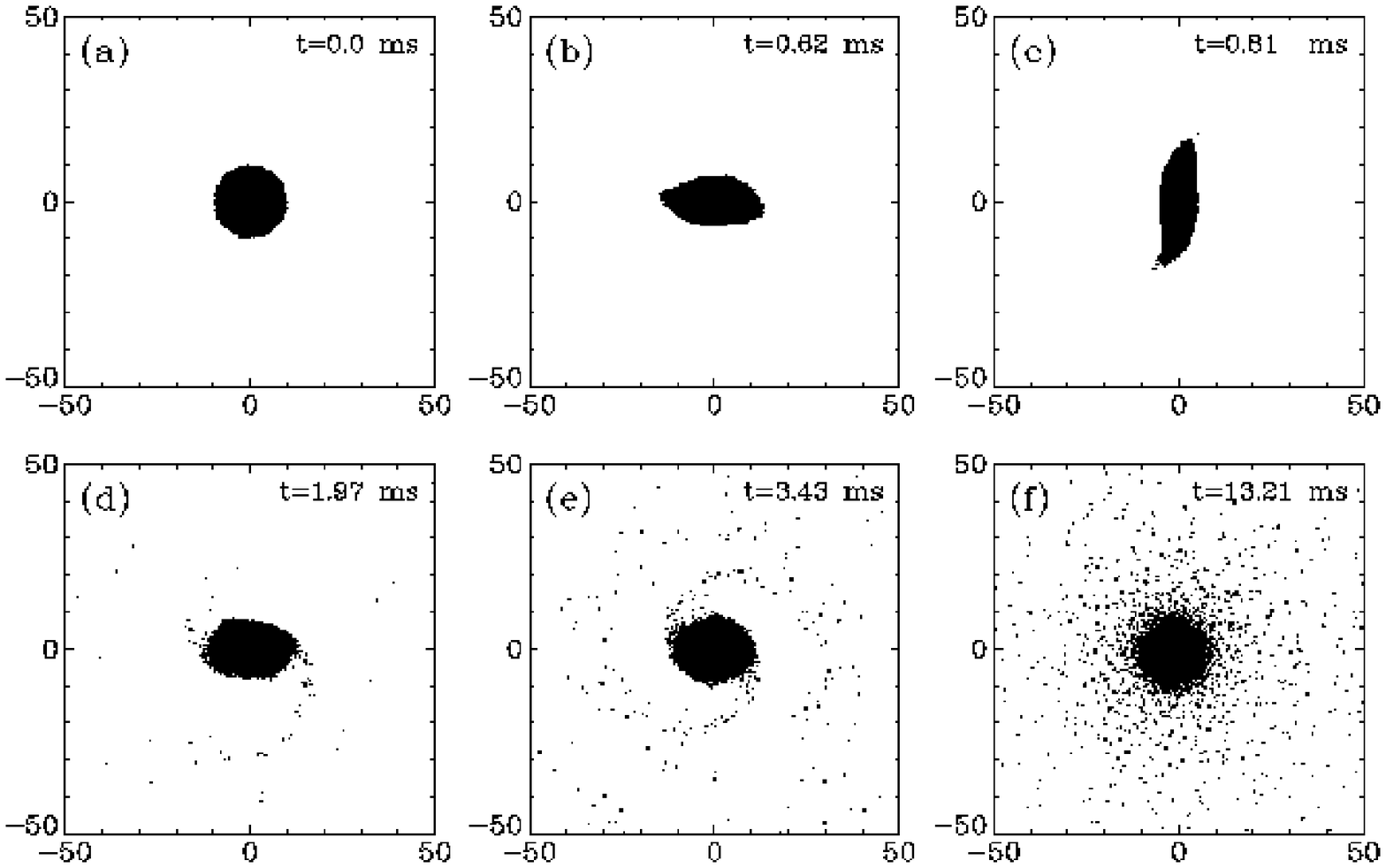}
\end{center}
(B)
\begin{center}
\epsfxsize=13cm \leavevmode \epsfbox{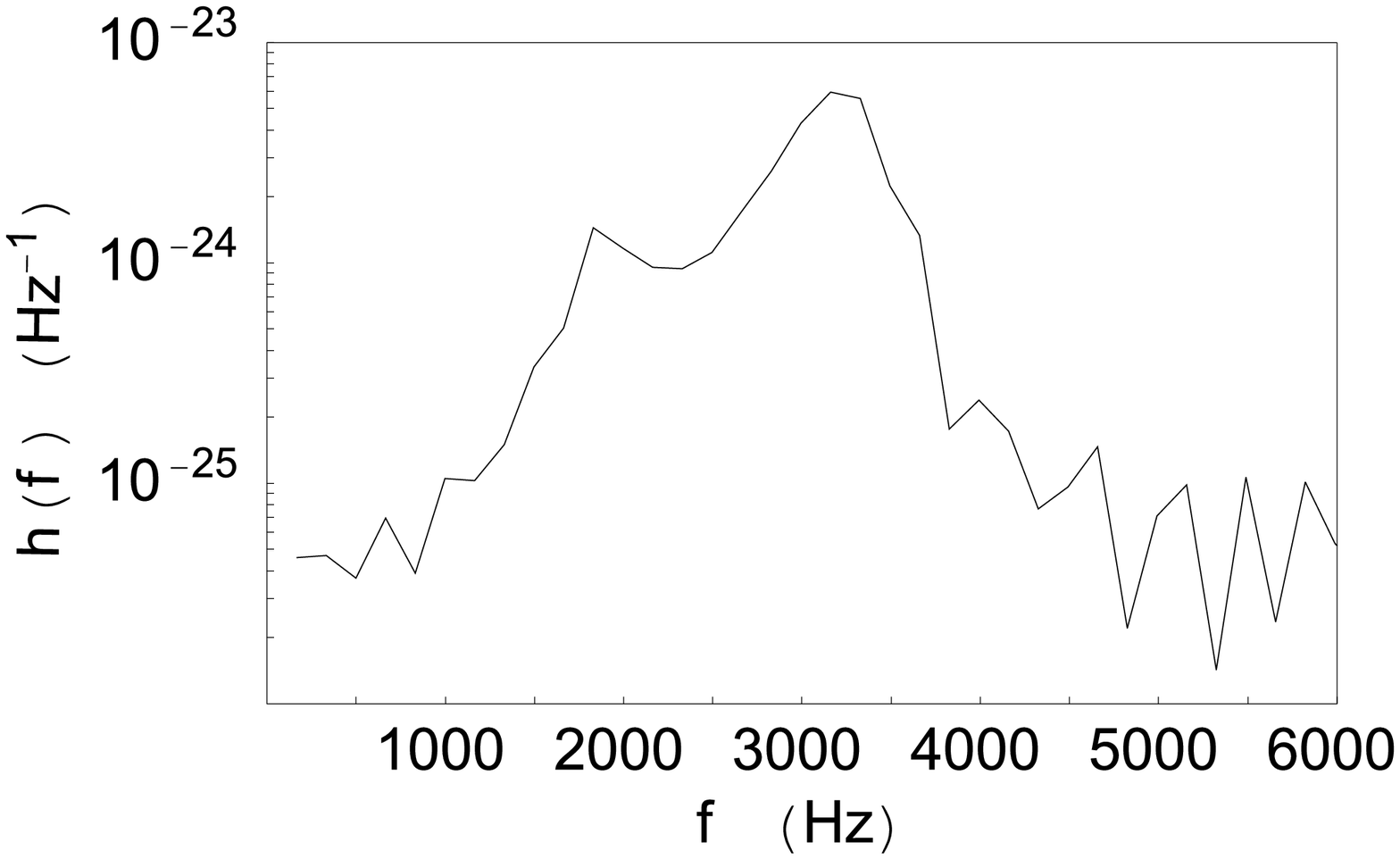}
\end{center}
\label{fig:sourcerrs}
\end{figure}

For the simulation presented here, the pre-collapsed stellar core is
modeled as a differentially rotating, compressible, axisymmetric fluid
which uses a polytropic equation of state. The stellar core is assumed
to have collapsed to the point where centrifugal hangup occurs,
reaching near neutron star densities (polytropic index $n=0.5$) prior
to the onset of a global dynamical instability.  To maximize
relativistic effects, the core is assumed to have a total mass of $M_t
= 1.4 M_\odot$, and equatorial radius, $R_{\rm eq} = 10\;{\rm km}$.
However, the collapse phase itself is not simulated.  The star rotates
counterclockwise about the $z-$axis at at equatorial surface speed of
approximately $0.122 {\rm c}$ \cite{Houser2000,Houser00a}. The event
was modeled as occurring at a distance of 1~Mpc as an optimistic
estimate of the distance necessary for antennas to see roughly one
event per year~\cite{Thorne95}.  A series of snapshots of
the evolution along with a graph of the waveform
used is shown in Figure~\ref{fig:sourcerrs}.

Strongly magnetized neutron stars are expected to form at the end
of Type II Supernova collapse.  For sufficiently strong fields,
misalignment between the rotation and magnetic axes can distort
the star by trapping the ejecta within the field \cite{BG96}.
This can cause a reduction in the angular momentum in a rapidly
rotating core through magnetic braking, which can remove several
orders of magnitude from the initial angular momentum over long
enough time scales \cite{FH99}.  The inclusion of
magnetohydrodynamical effects into the existing numerical models
will have significant consequences on the stability and
subsequent evolution of the post-collapsed object \cite{Houser01}.

\section{Calculations and Results}

We used Eq.~\ref{eqn:snr} to find signal-to-noise ratios for both
spherical resonant mass detectors and the advanced LIGO interferometer
with RSE detecting the binary neutron star and the rapidly rotating
stellar core sources.  For each source, we calculated seventeen
signal-to-noise ratios; one each for a sphere and interferometer
configuration at eight different frequencies plus one for advanced LIGO
in a broadband mode.  Although signal-to-noise ratio is ultimately the
figure of merit for a gravitational wave antenna, the comparison of 
strain spectra (as in Fig.~\ref{fig:strain}) gives a full understanding 
of the relative merits of the two detectors.  We present the signal-to-noise
ratio calculations to show how each antenna performs astronomically.
Comparisons between the two instruments can be done solely on the basis
of noise performance and cross section.

The frequencies were set by the choice of diameter for the spheres. The
largest sphere, 3.25~m in diameter, has a mass of 50 tonnes. The
smallest sphere chosen has a diameter of 1.05~m. Any smaller, and the
cross section for gravitational wave detection (implicit in
Eq.~\ref{eqn:comsig}) becomes to small for any realistic sources.  The
resonant frequencies of these spheres are given by~\cite{Merkowitz}
\begin{equation}
f_0 = 1.62/(\pi d_{\mathrm{sph}}) \sqrt{(Y/\rho)},
\end{equation}
where $d_{\mathrm{sph}}$ is the diameter of the sphere, $Y$ is the
Young's modulus of the sphere material, and $\rho$ is the density of
the sphere material.  This choice of diameters, then, allows for
sensitivity between 795~Hz and 2461~Hz, see Table~\ref{table:phase}.

The most sensitive frequency of the interferometer's noise
spectrum was adjusted by changing the position of the signal
recycling mirror.  A change in position less than the wavelength
of the laser light results in a change in phase $\delta$ in
Eqs.~(\ref{g01}) and~(\ref{g02}).  This, then, changes the
frequency characteristic of the shot noise.

The appropriate $\delta$ was found from setting the derivative of the
shot noise with respect to frequency at the resonance frequency of the
sphere equal to zero.  This insures the minimum of the shot noise,
which is the dominant noise source at frequencies above 500~Hz, will be
at the same frequency as the sphere's most sensitive point.  Since the
frequency dependence of the shot noise is all in the function
\begin{equation}
|G_0\left(f\right)| = 1/\left(1/|G_{0,1}\left(f\right)| +
1/|G_{0,2}\left(f\right)|\right),
\end{equation}
this condition can be written
\begin{eqnarray}
0 & = & \partial |G_0\left(f\right)| \left/ \partial \delta \right. \nonumber \\
& = & \left(r_3 \left( r_1 \left(1+r_{2}^{2}\right) \sin\left(2
\pi f \tau_s + \delta\right) - r_2 \left( r_{1}^{2} \sin\left(2
\pi f \left( \tau_s - \tau_a\right) + \delta) +
 \sin\left(2 \pi f \left(\tau_a+\tau_s\right) + \delta
 \right) \right) \right) \right)
 \right. \nonumber \\ & & \left/  \left( 2 \left(1 + r_{1}^{2} r_{2}^{2} +
r_{1}^{2} r_{3}^{2} + r_{2}^{2} r_{3}^{2} -2 r_1 r_2
\left(1+r_{3}^{2}\right) \cos\left(2 \pi f \tau_a\right) - 2 r_1
\left(1 + r_{2}^{2}\right)r_3\cos\left(2
\pi f \tau_s + \delta\right)  \right. \right. \right. \nonumber \\
& & \left. \left. + 2 r_{1}^{2} r_2 r_3 \cos\left(2 \pi f \left(
\tau_s - \tau_a\right) + \delta\right) + 2 r_2 r_3 \cos\left(2
\pi f \left( \tau_a + \tau_s\right) + \delta\right))^{1/2}
\right) \right.
.
\end{eqnarray}
Using this equation, the appropriate phase shifts for the signal
recycling cavity were found, and are presented in
Table~\ref{table:phase}.

The bandwidth of the sphere is determined by the impedance matching
between the sphere and the SQUID amplifier.  With a three stage
transducer, this bandwidth is given by
\begin{equation}
\delta_{\mathrm{BW}} = \sqrt{m_2/m_1},
\end{equation}
where $\delta_{\mathrm{BW}}$ is the fractional bandwidth of the
sphere in the lossless limit~\cite{Price}.  For the choices of
masses in the transducer from Table~\ref{table:sphere}, this
bandwidth becomes
\begin{equation}
\delta_{\mathrm{BW}} = 10 \%.
\end{equation}

In the interferometer, the bandwidth of the high frequency
response is determined by the reflectivities of the input mirror
and the signal recycling mirror, $r_1$ and $r_3$ respectively. The
bandwidth of both the sphere and of the interferometer was
calculated from
\begin{equation}
\delta_{\mathrm{BW}} = 1/\left(2 f_0\right) \left(\int
S_{\mathrm{tot}} \left(f\right) \mathrm{d}f \right)^2 \left/ \int
S_{\mathrm{tot}}^2 \left(f\right) \mathrm{d}f. \right.
\end{equation}
To adjust the bandwidth of LIGO to better approximate the noise
spectrum of a sphere, the values $t_1$ and $t_3$ were then chosen to
get the minimum bandwidth possible.  The values used for all
frequencies were $t_1=\sqrt{0.03}$ and $t_3=\sqrt{0.005}$. The
bandwidth turns out to be higher than $10\%$, the sphere's bandwidth,
for all peak frequencies. Decreasing $t_3$ leads to losses dominating
over the transmittance which limits the peak sensitivity.

\begin{figure}
\caption{Energy signal-to-noise ratios for binary neutron star separated into an
inspiral and coalescence phase at 15 Mpc.  The waveform was divided at the 
dynamical instability frequency, when the neutron stars are about 20~km
apart. This source was simulated interacting with spherical resonant mass
antennas (shown with asterisks) and
interferometers operating with resonant sideband extraction (shown with
circles) and a broadband interferometer (shown with a dotted line). \\
(A) The binary neutron star inspiral phase.\\
(B) The binary neutron star coalescence phase.\\
}\vspace{-2em}
\hspace{-1.5em}
(A)
\begin{center}
\epsfxsize=11cm \leavevmode \epsfbox{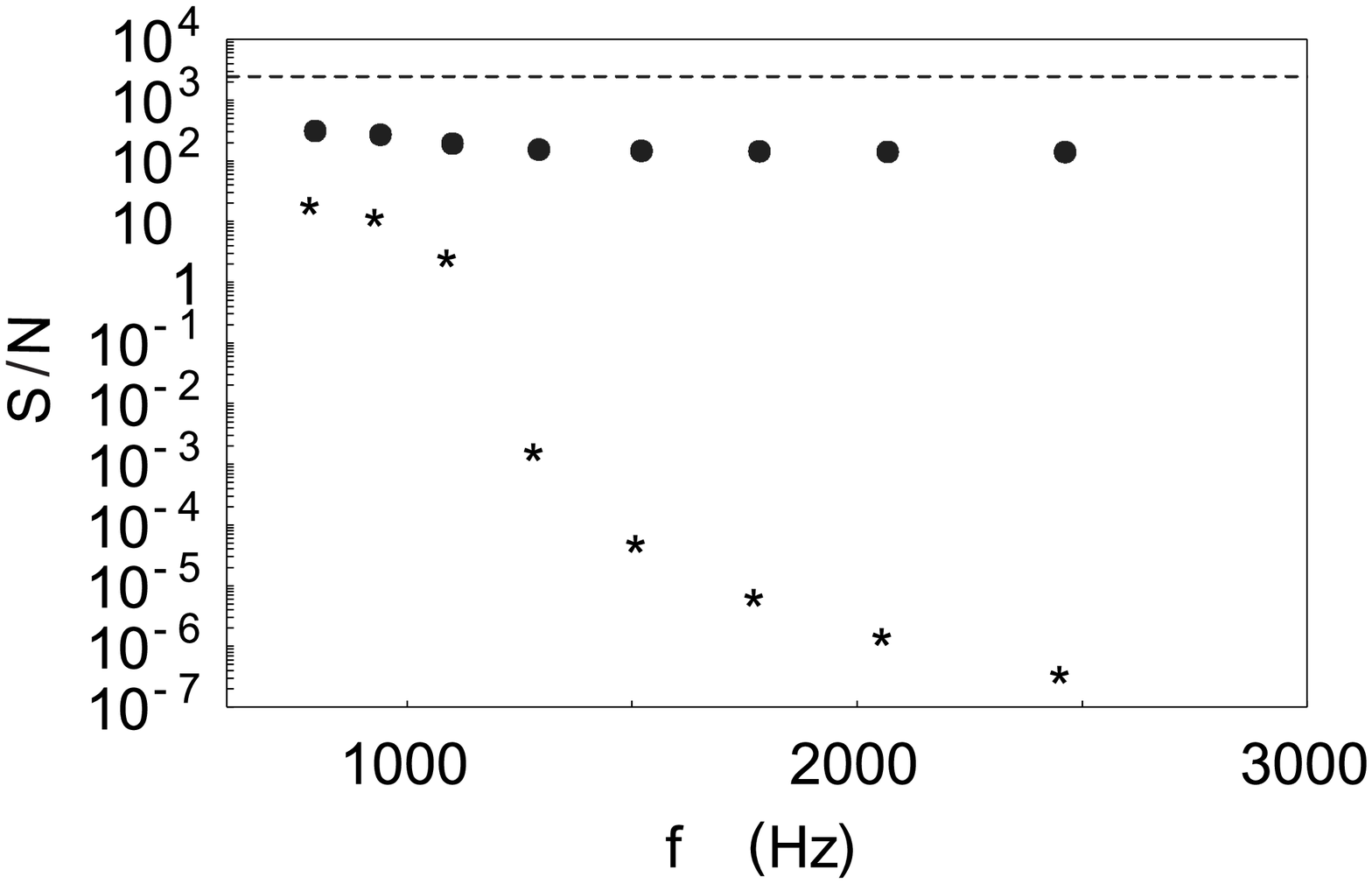}
\end{center}
(B)
\begin{center}
\epsfxsize=11cm \leavevmode \epsfbox{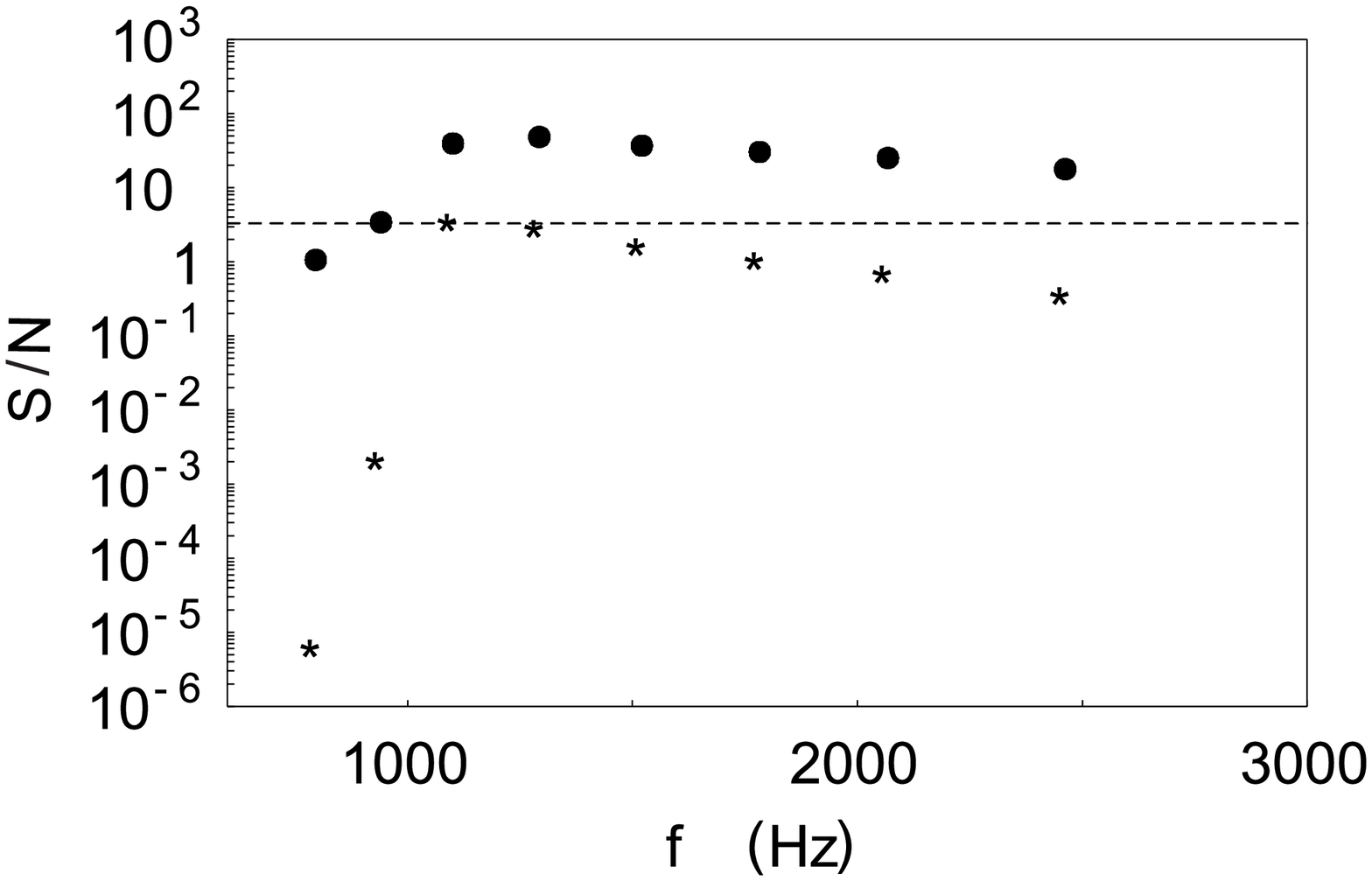}
\end{center}
\label{fig:separatesnr}
\end{figure}

The resulting signal-to-noise ratios for the spheres and interferometer
configurations are shown in Fig.~\ref{fig:bnssnr} for the binary
neutron star inspiral and coalescence, in Fig.~\ref{fig:separatesnr} for 
the inspiral and coalescence phases separately, and in Fig.~\ref{fig:rrssnr} for
the rapidly rotating stellar core undergoing a dynamical instability.
We also calculated signal-to-noise ratios for the sources interacting
with LIGO in a broadband configuration optimized for binary neutron star
inpiral. This involves changing the
input transmittance, $t_{1}^{2}$ to 0.005, the signal recycling mirror
transmittance, $t_{3}^{2}$ to 0.05, and the accumulated phase $\delta$
to 0.09.  These signal-to-noise ratios are shown in the same figures
with a dotted line.

\section{Conclusions}

\begin{figure}
\caption{Energy signal-to-noise ratios for binary neutron star
inspiral and coalescence.  This source was simulated interacting with
spherical resonant mass antennas (shown with asterisks) and
interferometers operating with resonant sideband extraction (shown with
circles) and a broadband interferometer (shown with a dotted line). \\
(A) The binary neutron stars were assumed at a distance of 15~Mpc. This
distance is the closest estimated for a single event a year.
Both detectors have high enough SNR's to reach this distance.\\
(B) The binary neutron stars were assumed at a distance of 200~Mpc.
This distance is enough for multiple events for year, and advanced LIGO
with RSE does have a SNR high enough to reach this distance.
}\hspace{-1.5em}
(A)
\begin{center}
\epsfxsize=10cm \leavevmode \epsfbox{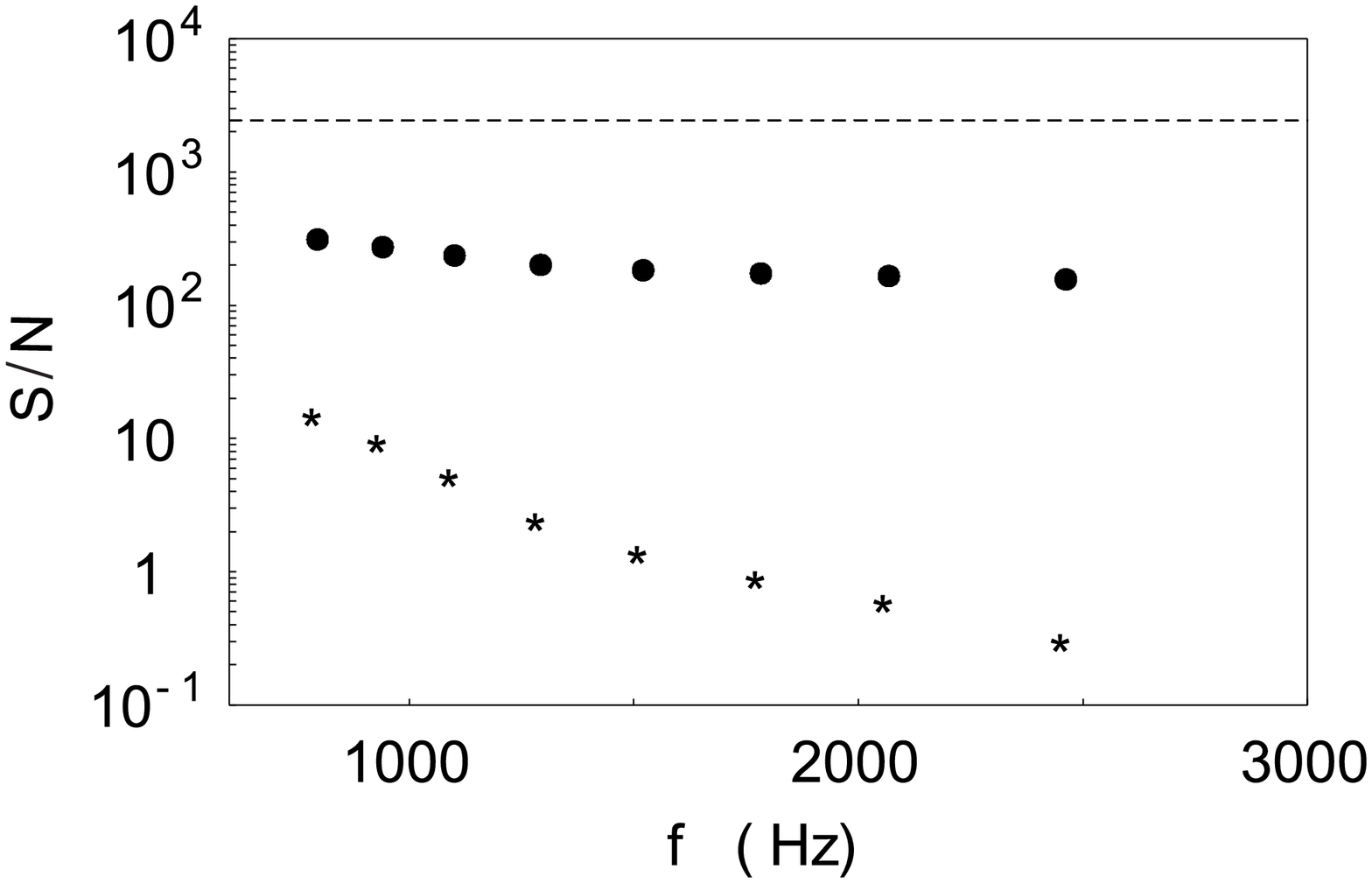}
\end{center}
(B)
\begin{center}
\epsfxsize=10cm \leavevmode \epsfbox{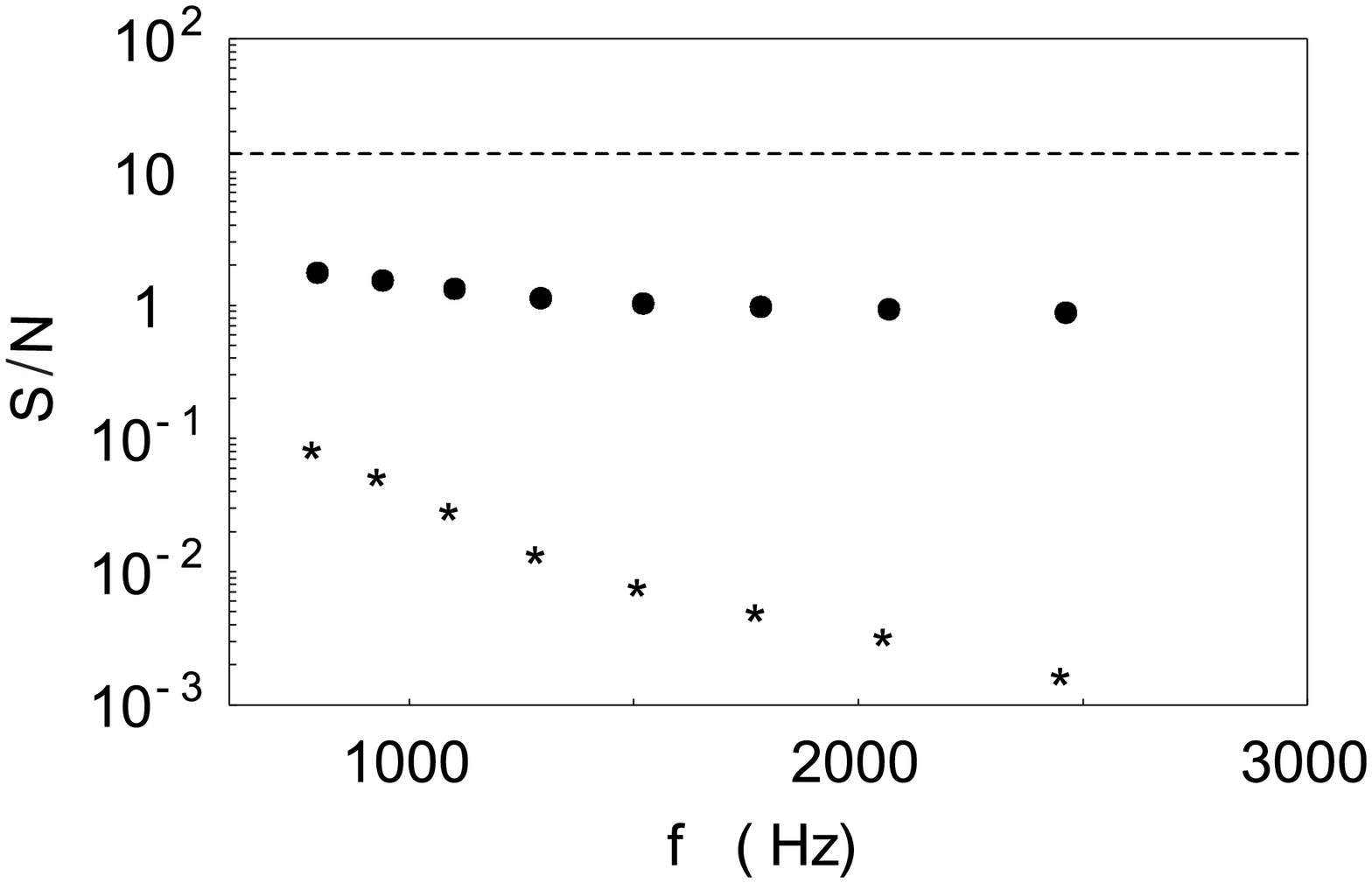}
\end{center}
\label{fig:bnssnr}
\end{figure}

\begin{figure}
\caption{Energy signal-to-noise ratios for a rapidly rotating stellar core
undergoing a dynamical instability.  This source was simulated
interacting with spherical resonant mass antennas (shown with
asterisks) and interferometers operating with resonant sideband
extraction (shown with circles) . The rapidly rotating core event was
assumed at a distance of 1~Mpc.}\label{fig:rrssnr}
\begin{center}
\epsfxsize=16cm \leavevmode \epsfbox{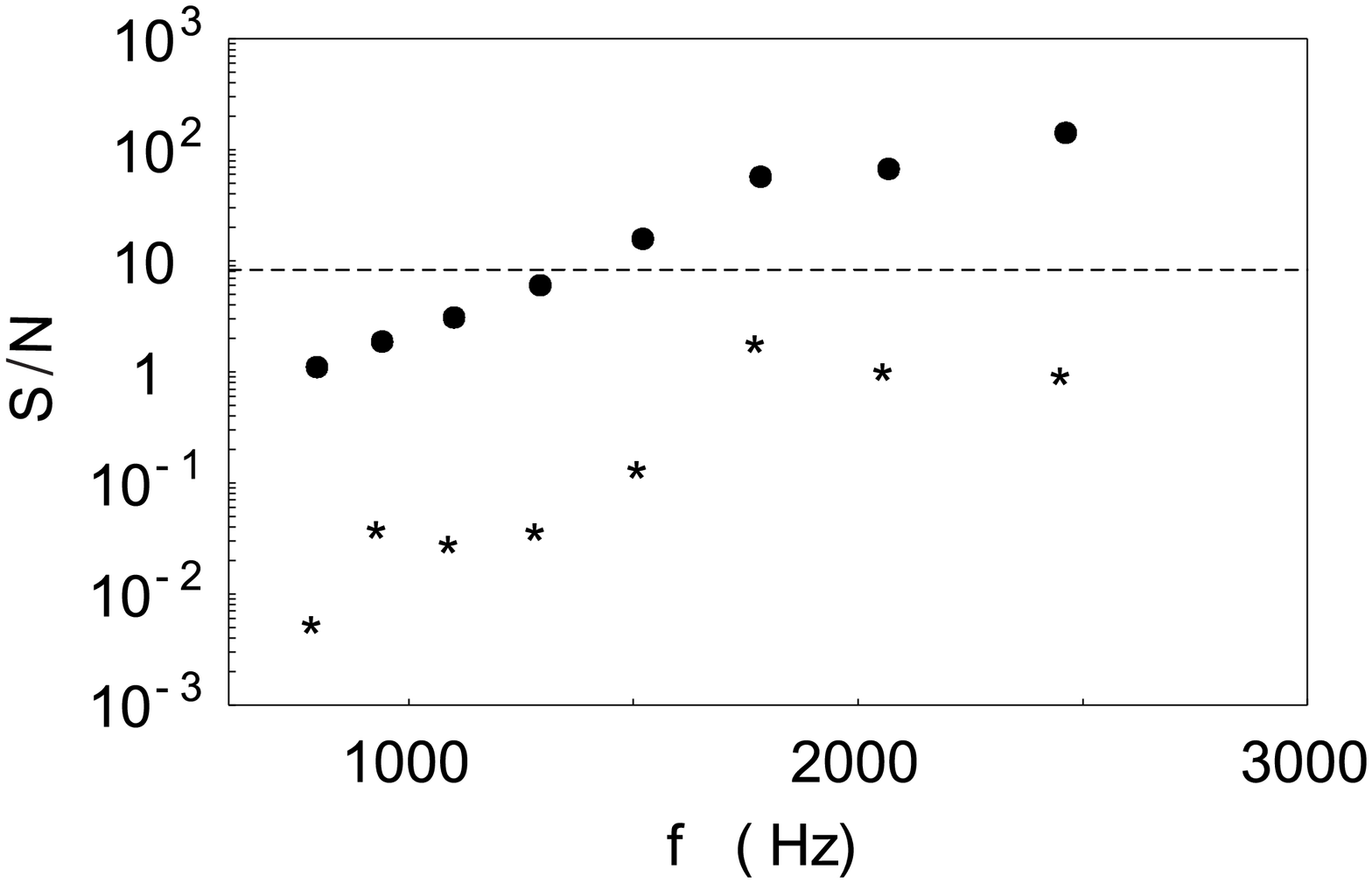}
\end{center}
\end{figure}

Interferometers utilizing resonant sideband extraction can be more
sensitive than the most sensitive spheres, both on and off resonance.
This condition remains true even when the effects of random
polarization and direction of the gravitational wave are accounted for.
Spheres are always optimally oriented towards incoming waves.
Figures~\ref{fig:bnssnr} and~\ref{fig:rrssnr} indicate that this
greater sensitivity translates into significantly higher SNR's for the
interferometer over spheres for the two sources we considered.

These two figures show how sensitive each technology is to the two
sources. A properly sized sphere can detect the inspiral signal of a
binary neutron star system at a distance of 15~Mpc.  This is far enough
to reach the nearer sections of the Virgo cluster of galaxies.
According to optimistics estimates~\cite{Harry,P91} this may be enough
to detect one event per year.  Advanced LIGO can see binary neutron
star events out to 200~Mpc with a single detector, the most likely
distance necessary to see multiple events per year~\cite{P91}. Advanced
LIGO will also be able to see rapidly rotating stellar core events at a
far enough distance to detect multiple events a year. Depending on the
size of the sphere, resonant mass technology may also have enough
sensitivity to see one or more rapidly rotating stellar core events a
year as well.

Detecting the coalescence phase of the binary neutron star event would 
provide information about the structure of these stars (e.g., the 
equation of state and the equatorial radius).  Advanced LIGO properly
tuned to a high frequency, narrowband configuration provides the highest
signal-to-noise ratio for this source.  In this mode, Advanced LIGO has
enough sensitivity to detect the coalescence waveform at a distance of 
100~Mpc.  This may be enough to actually see such an event during the 
expected lifetime of Advanced LIGO.    Advanced LIGO tuned to
1292~Hz, where the highest signal-to-noise ratio is obtained, is
mostly sensitive to the early stages of coalescence.  Choosing both a
sphere radius of 70~cm and a phase, $\delta$, for the interferometer of
0.04613, allows these antennas to be tuned to the 3700~Hz of the rotating
bar peak.  The SNR for the sphere at this frequency is only $4.5 \times
10^{-2}$ at  15~Mpc.  Advanced LIGO has a SNR of 6.39 at this distance,
but event rate predictions are pessimistic about a coalescence happening
this close.

A comparison of the signal-to-noise ratios found in the previous
paper~\cite{Harry} for binary neutron star events with
Figures~\ref{fig:bnssnr} and~\ref{fig:rrssnr} shows that the addition
of the gravitational wave back reaction to the model does change the
waveform of the coalescence phase of the binary neutron star evolution. 
It is important for deciding the best configuration of advanced LIGO to
know the details of the coalescence waveform.  Other effects, notably
inclusion of the magnetic fields in the
neutron stars and post-Newtonian corrections~\cite{Rasio}, may change
all these waveforms, especially for the coalescence and the rapidly
rotating stellar core events.  We have used the best available models to
predict the gravitational radiation but further improvements are probable
and our results should be seen in this light.

Despite the sensitivity advantages of interferometers, spheres do have
benefits which should allow them to find a niche in the field of
gravitational wave detection. Having simultaneous detection of a single
event by two completely different technologies will help confirm
signals with marginal SNR's; a near certainty in the early years of
gravitational wave astronomy. Having a sphere near to an interferometer
site will also allow for correlated searches for stochastic backgrounds
of cosmological gravitational waves~\cite{maggiore,vitale}.  It is
conceivable that such pairing may occur in Louisiana between LIGO and
Louisiana State University and in Italy between Virgo and an Italian
sphere.

Spheres may be particularly well suited for detecting scalar
radiation~\cite{coccia2,maggiore2,bianchi} because of their
symmetry properties. This would allow for exploration of gravity
beyond the predictions of general relativity.  The comparatively
low cost of spherical antennas in relation to interferometers
could allow for construction of more individual detectors which
are located more widely around the globe. The decades of
experience working with bar detectors will provide useful
background for sphere projects.  Operation in conjunction with the
interferometer network, an array of spherical detectors will be a
valuable asset to the worldwide effort to develop gravitational
wave astronomy.

\section{Acknowledgments}

We would like to thank Peter~R.~Saulson for his support, advice, and comments;
Valerie~Williams for careful editing, Andri~M.~Gretarsson and Steve Penn
for careful reading and comments. One of us (GMH) would like to thank
the MIT LIGO group for support while finishing the manuscript.  The LIGO
Visitors Program was instrumental in completing this work; we thank Syd
Meshkov, Barry Barish, Gary Sanders, and Rai Weiss for their help.  We
also thank Rai Weiss for helpful comments on the sources, Kip Thorne
for help and inspiration, as well as Gabriela Gonzalez, David Tanner, and
Gary Sanders for careful reading and comments. The sources were modeled on 
computers at the
Harvard-Smithsonian Center for Astrophysics and MIT. We especially thank
L. Sam Finn for making {\tt BENCH} available to us as well as everyone who
contributed to its development. This work was supported by Syracuse
University, U.S. National Science Foundation Grant No. PHY-9900775 and
9603177, the University of Glasgow, and PPARC.

\begin{table}
\caption{Parameters used in model of spherical, resonant mass,
gravitational wave antennas.}\label{table:sphere}
\begin{tabular}{llrl}
\multicolumn{1}{l}{Parameter} & \multicolumn{1}{c}{Name} &
\multicolumn{1}{l}{Value} &
    Source \\
$Q_{\mathrm{s}}$ & Sphere quality factor, Al & $40 \times 10^6$ & \cite{Duffy90} \\
$Q_{1}$ & Intermediate mass quality factor, Al & $40 \times 10^6$ & \cite{Duffy90} \\
$Q_{2}$ & Transducer mass quality factor, Nb & $40 \times 10^6$ & \cite{Duffy94,Geng} \\
$T$ & Temperature & 50~mK & \cite{Coccia} \\
$T_{n}$ & SQUID noise number & 1 & \cite{Harry2000} \\
$d_{\mathrm{c}}$ & Sensing coil diameter & 9~cm & \cite{Harry,thesis} \\
$\kappa$ & Electrical spring constant per area & $3.78 \times
10^8$~N/m$^{3}$ & \cite{Harry} \\
$m_{\mathrm{s}}/m_{1}$ & Mass ratio & 100 &
\cite{Harry} \\
$m_{1}/m_{2}$ & Mass ratio & 100 & \cite{Harry}
\end{tabular}
\end{table}

\begin{table}
\caption{Parameters used for model of interferometric
gravitational wave detector.}\label{table:ligo}
\begin{tabular}{llrl}
\multicolumn{1}{l}{Parameter} & \multicolumn{1}{c}{Name} &
\multicolumn{1}{l}{Value} &
    Source \\
$L$ & Interferometer arm length & 4000~m & \cite{ligoII} \\
$L_{\mathrm{rec}}$ & Recycling cavity length & 10~m &\\
$f_{\mathrm{seismic}}$ & Seismic noise cutoff & 10~Hz &
\cite{ligoII,Giame,Bertolini}\\
$\lambda$ & Wavelength of laser light & 1.064 $\mu$m &
\cite{ligoII} \\
$P$ & Laser Power & 125~W & \cite{ligoII,Byer} \\
$\eta$ & Photodiode quantum efficiency & 0.9 & \cite{stanford} \\
$w_{1}$ & Gaussian width of laser at input mirror & 6~cm &\\
$w_{2}$ & Gaussian width of laser at end mirror & 6~cm &\\
$\alpha$ & Relative power loss in beamsplitter & $3.5 \times
10^{-3}$ &  \\
$\beta$ & Relative power loss at each mirror & $3.75 \times
10^{-5}$ & \\
$a_{\mathrm{coat}}$ & Relative absorption of coating at input mass
& $1 \times 10^{-6}$ &\\
$A_{g}$ & Beamsplitter material absorption coefficient & $40
\times 10^{-4}$~m$^{-1}$&\\
$t_{\mathrm{BS}}$ & Thickness of beamsplitter & 12~cm & \\
$t_{1}^2$ & Power transmittance of input mirror & 0.03 & \\
$t_{3}^2$ & Power transmittance of signal recycling mirror & 0.005 \\
$r$ & Radius of mirrors & 14~cm & \\
$\ell$ & Thickness of mirrors & 12~cm &  \\
$\phi$ & Mirror material loss angle & $5.0 \times 10^{-9}$ &
\cite{Braginsky} \\
$L_{\mathrm{sus}}$ & Length of suspension & $0.588$~m &  \\
$d$ & Total ribbon thickness & 1.7~mm & \\
$\phi_{\mathrm{bulk}}$ & Loss angle for the bulk ribbon material
& $3.3 \times 10^{-8}$ & \cite{Gretarsson} \\
$d_{s}$ & Dissipation depth of ribbon material & $182~\mu$m &
\cite{Gretarsson} \\

\end{tabular}
\end{table}
\begin{table}
\caption{Parameters of the signal recycling mirror to simulate the
frequency response of spheres.  The transmittance of the input mirrors
was held constant at $t_1^2 = 0.03$.  The transmittance of the signal
recycling mirror was held constant at $t_3^2 = 0.005$. Note that the
resonance frequency of the 1.25~m sphere in~\protect\cite{Harry} was a
typographical error, the value listed here is
correct.}\label{table:phase}
\begin{tabular}{cccl}
\multicolumn{1}{c}{Diameter $d_{\mathrm{sph}}$}(m) &
\multicolumn{1}{c}{Frequency $f_0$ (Hz)} &
\multicolumn{1}{c}{Bandwidth $\Delta f/f_0$} &
\multicolumn{1}{l}{Phase $\delta$}  \\
3.25 & \,795 & 0.170 & 0.2271 \\
2.75 & \,940 & 0.172 & 0.1921 \\
2.35 & 1100  & 0.182 & 0.1641 \\
2.00 & 1292  & 0.200 & 0.1395 \\
1.70 & 1520  & 0.225 & 0.1182 \\
1.45 & 1782  & 0.254 & 0.1005 \\
1.25 & 2067  & 0.290 & 0.08619 \\
1.05 & 2461  & 0.330 & 0.07179
\end{tabular}
\end{table}


\pagebreak[4]

\pagebreak[4]

\pagebreak[4]

\pagebreak[4]

\pagebreak[4]

\pagebreak[4]

\pagebreak[4]

\pagebreak[4]

\pagebreak[4]

\pagebreak[4]

\end{document}